\documentstyle[11pt,psfig,rotating]{article}
\begin{document}

\today

\begin{center}

{\LARGE\bf  Spectroscopic studies  in open quantum systems \\}

\vspace*{.6cm}

{\large 

{\large {\bf I.~Rotter$^*$, 
E.~Persson$^+$,
K.~Pichugin$^\dagger$, 
and 
P.~${\rm \check S}$eba$^{\dagger \ddagger}$
}}\\[0pt]

\vspace{0.5cm}

{\it 
$^*$Max-Planck-Institut f\"ur Physik komplexer Systeme, D-01187
Dresden, Germany \\[0pt]
$^+$Institut f\"ur Theoretische Physik,
Technische Universit\"at Wien,  A-1040 Wien,  \"Osterreich \\[0pt] 
$^\dagger$ Institute of Physics, Czech Academy of Sciences,
Cukrovarnicka 10, Prague, Czech Republic \\[0pt]
$^\ddagger$Department of Physics, Pedagogical University, Hradec Kralove,
Czech Republic\\
}
}

\end{center}

\vspace*{.5cm}

\begin{abstract}
The  spectroscopic properties of an open quantum system
are determined by the eigenvalues and eigenfunctions of an effective
Hamiltonian ${\cal H}$ consisting of the Hamiltonian ${\cal H}_0$ 
of the corresponding closed system and a non-Hermitian  correction term $W$
arising from the interaction via the continuum of decay channels.
The hermitian part of  ${\cal H}$ is
${\cal H}_0 + \Re(W)$ while the anti-hermitian part 
is $\Im(W)$. 
This holds for a many-particle system as well as for a microwave resonator.
The eigenvalues ${\cal E}_R$ of ${\cal H}$ are complex. They are the poles of
the $S$-matrix  and provide both the
energies and widths of the states. 
We illustrate the
interplay between $\Re({\cal H})$ and $\Im({\cal H})$
by means of the
 different interference phenomena between two neighboured resonance states.
Level repulsion along the real axis appears if the 
interaction is caused mainly by (the non-diagonal part of)
$\Re({\cal H})$   while
a bifurcation of the widths appears if the 
interaction occurs mainly due to (the non-diagonal part of) $\Im({\cal H})$. 
We then calculate 
 the poles of the $S$-matrix and the corresponding wavefunctions for a
rectangular microwave resonator with a scatter
  as a function of the area of the
resonator as well as of the degree of opening to a guide. 
The calculations are performed by using the method of exterior complex 
scaling. $\Re(W)$ and $\Im(W)$
cause  changes in the structure of the wavefunctions
which are permanent, as a rule.
At full opening to the lead, short-lived
collective states are formed together with long-lived trapped
states. The wavefunctions of the short-lived states at full opening to the
lead are very different from those at small opening.
The resonance picture obtained from the microwave resonator shows all the
characteristic features known from the study of many-body systems 
{\it in spite of} the absence of two-body forces. 
The poles of the $S$-matrix determine  the conductance of the resonator.
Effects arising from the interplay between resonance trapping and level 
repulsion along the real axis are
not involved in the statistical theory (random matrix theory).

\end{abstract}

\vspace{.5cm}

\section{Introduction}

Since more than ten years, interference phenomena in open quantum systems are 
studied theoretically in the framework of different models. Common to all
these studies is the appearance of different time scales as soon as the
resonance states start to overlap (see
\cite{nucl1,nucl2,marost,chem,general,hemuro,pepirose,seromupepi}
and  references in these papers to older ones).
  Some of the states align with
the decay channels and become short-lived while the remaining ones decouple
to a great deal from the continuum of decay channels and become long-lived.
The wavefunctions show  permanent changes: they are mixed 
strongly in the basic
  wavefunctions of the corresponding closed system. 

In   many-body systems the interaction is caused, above all, by 
two-body forces between the constituents of the system.
The additional interaction connected with avoided level crossings 
is believed, usually, to lead 
only to an exchange of the wavefunctions
but not to permanent changes of their structure. This conclusion results
from many spectroscopic studies on closed systems with discrete states.
Recent investigations 
in the framework of a schematical model \cite{nucl1} showed however that, 
in the case of collective resonance states, 
permanent changes in the wavefunctions occur 
due to the interplay between the
real and imaginary parts of the different coupling matrix elements.

The mixing of the resonance states of a microwave resonator is not caused by
two-body forces. A mixing of the states may occur 
only as a result of avoided level crossings. It is therefore an interesting 
question whether or not some permanent mixing in the wavefunctions 
of a microwave cavity can arise.
In \cite{reichl2}, the changes in the structure of the wavefunctions at
avoided crossings in a strongly driven (closed)
square potential well system are 
studied. The avoided crossings are shown to lead, in some cases, 
to  temporary and, in other cases, to permanent 
changes  as a function of driving field strength. 

The avoided level crossings
are related to exceptional points in the complex plane \cite{hemuro}. The
coupling induced by avoided level crossings is therefore surely connected
with the coupling matrix elements of the discrete states of the closed system
to the continuum. 
In the continuum shell model, these coupling matrix elements are  complex 
\cite{drozdz}. Interferences appear when more than one channel are open. 
It is possible therefore that interferences of different type 
may provide eventually
a permanent mixing of the wavefunctions of the resonance states.
 A similar study for microwave
cavities does not exist.

It is the aim of the present paper to study the resonance picture of an
open microwave resonator in detail. 
We show that its  characteristic features are the same as
those which are known  from  open  many-body quantum systems. 
This means that not only two-body forces  play a  role
for the interaction among the resonance states but also the interaction $W$ via
the continuum is important. 
Neither $\Re(W)$ nor $\Im(W)$ can be neglected, generally.
They are important near avoided crossings in the complex plane and
their interplay can not be neglected  when  $\Re(W)$ and $\Im(W)$
are of the same order of magnitude.  
As a result, permanent changes in the structure of the wavefunctions
appear, as a rule.
The basic assumptions of the statistical theory (random matrix theory) are
fulfilled only when  the interferences caused by
the hermitian and anti-hermitian parts of the Hamiltonian
 can be neglected to a good
approximation. This is the case  e.g. for a 
Gaussian orthogonal ensemble coupled weakly to the continuum.

In Section 2 of the present paper, 
the Hamiltonian of an open quantum system and the
relation of its eigenvalues to the poles of the $S$-matrix is considered.
The formalism can be applied to a many-body system as well as to a microwave
resonator. 
The Hamiltonian is  non-hermitian and the eigenvalues provide the energies as
well as the widths of the  states. In
Section 3, the avoided crossing of two resonance states is traced. The
differences between the mixing of the states  
due to the hermitian and the anti-hermitian part of ${\cal H}$ are the central
point of discussion.
The hermitian part causes  an equilibration of the states in relation to the
time scale which is accompanied
by level repulsion along the real axis (energy). In contrast to this, 
the anti-hermitian part
leads to an attraction of the levels in energy and to a bifurcation
of the widths (formation of different time scales).  
These processes are characteristic for the interplay among resonances which
takes place locally in more complicated systems \cite{mudiisro}.

In Section 4, the resonance structure of a rectangular microwave resonator
 coupled to one lead is studied. Inside the resonator is a circular scatter. 
Level repulsion in the complex plane appears. It can be seen sometimes as 
level repulsion along the real energy axis. In other cases,  
a bifurcation of the widths occurs. 
The changes in the structure of the wavefunctions
are permanent, as a rule. 
Collective states are formed at strong coupling
to the lead. The structure of their wavefunctions has almost nothing in common 
with the structure of the wavefunctions of these states at small coupling.
Together with the collective states, long-lived trapped states appear.
The conductance of the microwave resonator is studied
 after coupling it to a second lead. The conductance peaks are determined by 
the poles of the $S$-matrix, which move as a function of the 
coupling strength  between cavity and leads. 
The results are discussed in Section 5 and some conclusions are drawn 
in the last section.

\section{The Hamilton operator of an open quantum system}

The function space of an open quantum system consists of two parts: the
subspace of discrete states ($Q$-subspace)
  and the subspace of scattering states ($P$-subspace).  
The discrete states are the
states of the closed system which are embedded into the continuum of
scattering states. Due to the coupling of the discrete states to the
continuum, they can decay and get a finite lifetime.

Let us define  two sets of wavefunctions by
solving first the Schr\"odinger equation 
  $(H^{\rm cl} - E_R^{\rm cl})\; \Phi_R^{\rm cl} = 0$ 
for the discrete states  of the closed system
and secondly the Schr\"odinger equation 
  $ (H^{cc} - E^{(+)}) \; \xi_E^{c} = 0 $
for the scattering states of the environment.
Note that the closed system can be a many-particle quantum system or a 
system like a microwave resonator. The only condition is that it can be
described quantum mechanically by the hermitian Hamilton operator
$H^{\rm cl}$. In the case of the flat microwave resonator, this is possible 
by using the analogy to the Helmholtz equation.
Then, the     $Q$ and $P$ operators can be defined by 
\begin{eqnarray}
Q  =  \sum_{R=1}^N \;  | \Phi_R^{\rm cl} \rangle
      \langle \Phi_R^{\rm cl} | \;  \qquad
P  =  \sum_{c=1}^\Lambda \int_{\epsilon_c}^\infty \; dE \;
   \; | \xi_E^c \rangle
  \langle \xi_E^c |  
\label{eq:pqop} 
\end{eqnarray}
and    $Q \cdot \xi_E^{c} = 0 \, ;  \; 
 P \cdot \Phi_R^{\rm cl} = 0 $.  
In order to perform spectroscopic studies,
we do not use any statistical assumptions (for details see \cite{ro91}).

Assuming
  $ Q + P = 1$, we can determine a 
third wavefunction by solving
the  scattering problem
   $( H^{cc} - E^{(+)}) \; \omega_R = - \sum_c \gamma_{Rc} \; \xi_E^c$
with source term.  The source term is determined by the coupling 
matrix elements $\gamma_{Rc}$  between the two subspaces. 
 Further, we identify
$H^{\rm cl}$ with  ${\cal H}_0 \equiv QHQ$ 
where   $(H-E)\Psi = 0$ is the Schr\"odinger equation in the 
total function space $P+Q$.
Then,   the solution $ \Psi  =  P \Psi + Q \Psi $ 
in the total function space is \cite{ro91}
\begin{eqnarray} 
 \Psi 
  & = & \xi_E^c + \frac{1}{2 \pi} \sum_{R=1}^N \sum_{R'=1}^N
   (\Phi_R^{\rm cl}   +  \omega_R)    \cdot  
    \langle \Phi_R^{\rm cl} |\frac{1} {E - 
{\cal H}} | \Phi_{R'}^{\rm cl}\rangle \;
\gamma_{R'c}
\nonumber    \\
  & = & \xi_E^c + \sum_{R=1}^N
  {\tilde \Phi}_R   \cdot
  \frac{
\tilde\gamma_{Rc}}
  {E - {\tilde E}_R + \frac{i}{2} {\tilde \Gamma}_R}  \; .
\label{eq:wf1}
\end{eqnarray}
Here,
\begin{eqnarray}
 {\cal H}  =
{\cal H}_0  + W
\label{eq:heff}
\end{eqnarray}
is the effective Hamilton operator appearing in the $Q$-subspace due 
to the coupling to the continuum,
  ${\tilde \Phi}_R$  are the eigenfunctions of ${\cal H}$ and
  $\tilde {\cal E}_R \equiv {\tilde E}_R - \frac{i}{2} \;
 {\tilde \Gamma}_R$ its eigenvalues. They provide the wavefunctions,
energies and widths, respectively, of the resonance states.
The $\gamma_{Rc}$ are the coupling matrix elements between the discrete states
$\Phi_R^{\rm cl}$ and the continuum of scattering states $\xi_E^c$, while the
$\tilde \gamma_{Rc}$
are those between the resonance states $\tilde \Phi_R$ and the continuum.
The matrix elements of $W$ consist of the principal value integral
and the residuum \cite{ro91},
\begin{eqnarray}
 W_{R'R}^{\rm ex}
 & = & \frac{1}{2 \pi} \sum_{c=1}^\Lambda {\cal P} \int
\limits_{\epsilon_{c}}^{\infty} dE' \; 
\frac{\gamma_{Rc} \; \gamma_{R'c}}{E - E'} \; - \;
 \frac{i}{2} \sum_{c=1}^\Lambda 
 \gamma_{Rc} \;  \gamma_{R'c}
 \; .
\label{eq:extmi1}
\end{eqnarray}
Here, $c = 1, ... , \Lambda$ are the channels which open at the energies  
$\epsilon_{c}$. They describe the external mixing of two states via 
the continuum of decay channels. As a rule, both parts 
$\Re(W)$ and $\Im(W)$ are non-vanishing.   

Note, the expressions (\ref{eq:wf1}),  (\ref{eq:heff}) and
(\ref{eq:extmi1}) 
follow by formal  rewriting the Schr\"odinger equation $(H-E)\Psi = 0$
with the only 
condition that $Q$ and $P=1-Q$ are defined in such a manner
that the channel wavefunctions of the $P$ subspace are uncoupled              
\cite{ro91}. Otherwise, the eigenvalues and eigenfunctions of ${\cal H}$
have no physical meaning.
The  $\tilde E_R, \; \tilde \Gamma_R, \; \tilde\gamma_{Rc}$ and
 $\tilde \Phi_R$ are energy dependent functions,
generally. 

The resonance part of the $S$-matrix is \cite{ro91}
\begin{eqnarray}
S_{cc'}^{(\rm res)} = 
i \sum_{R=1}^N \frac{\tilde\gamma_{Rc'} 
\tilde\gamma_{Rc}}{E - {\tilde E}_R + \frac{i}{2} 
{\tilde \Gamma}_R} \; .
\label{eq:smares} 
 \end{eqnarray}
We underline that the 
$\tilde \gamma_{Rc}, ~\tilde E_R$,  $\tilde \Gamma_R$ and $\tilde \Phi_R$ 
are functions which are calculated inside the formalism. 
They contain the contributions of  $\Im(W)$ and of $\Re(W)$. The
$\tilde \gamma_{Rc}$ and $\tilde \Phi_R$ are complex.
 
\section{Avoided crossing of two resonance states}

\subsection{Schematical study}

\begin{figure}
\begin{minipage}[tl]{7.4cm}
\psfig{file=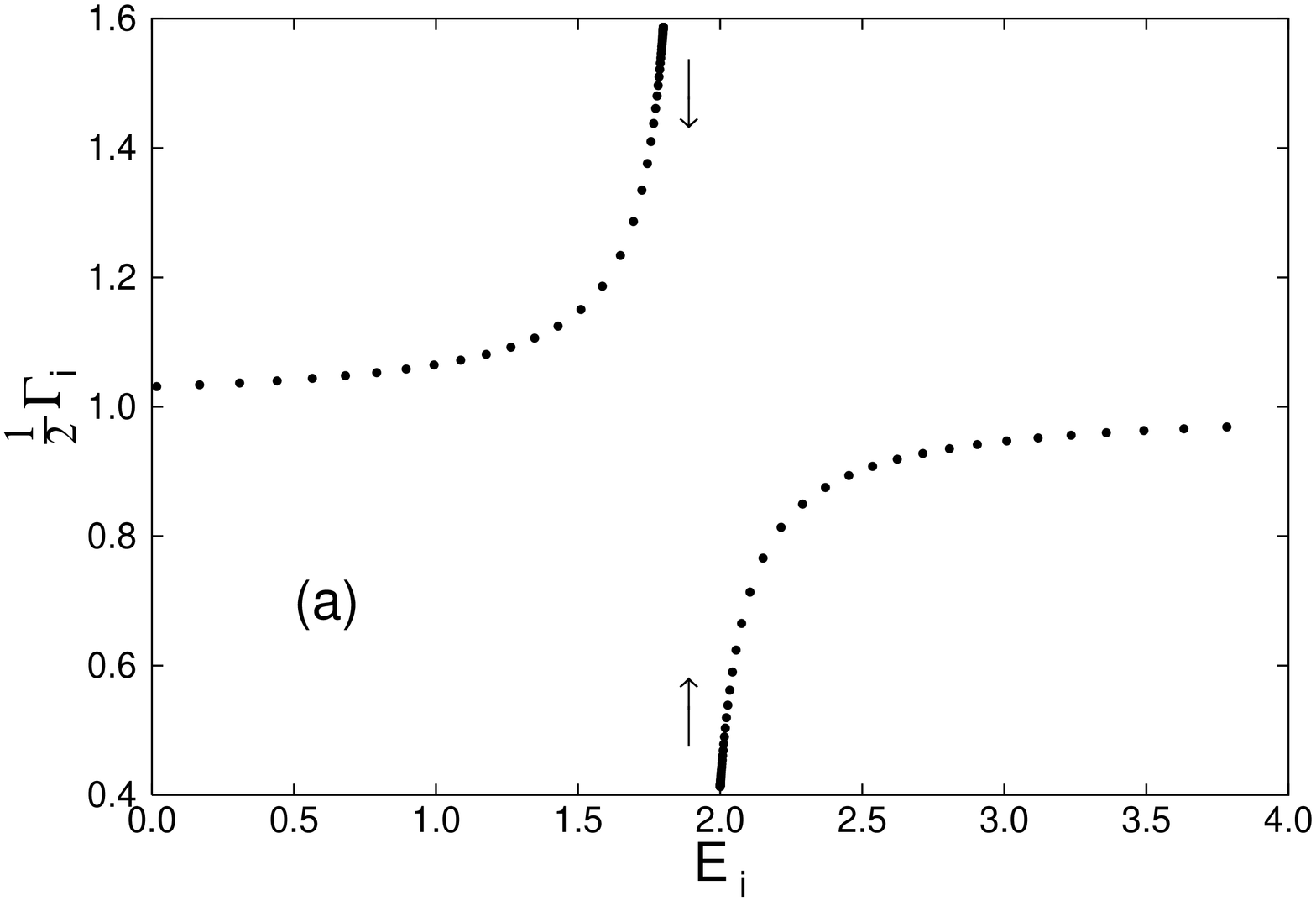,angle=-90,width=7.4cm}
\end{minipage}
\begin{minipage}[tr]{7.4cm}
\psfig{file=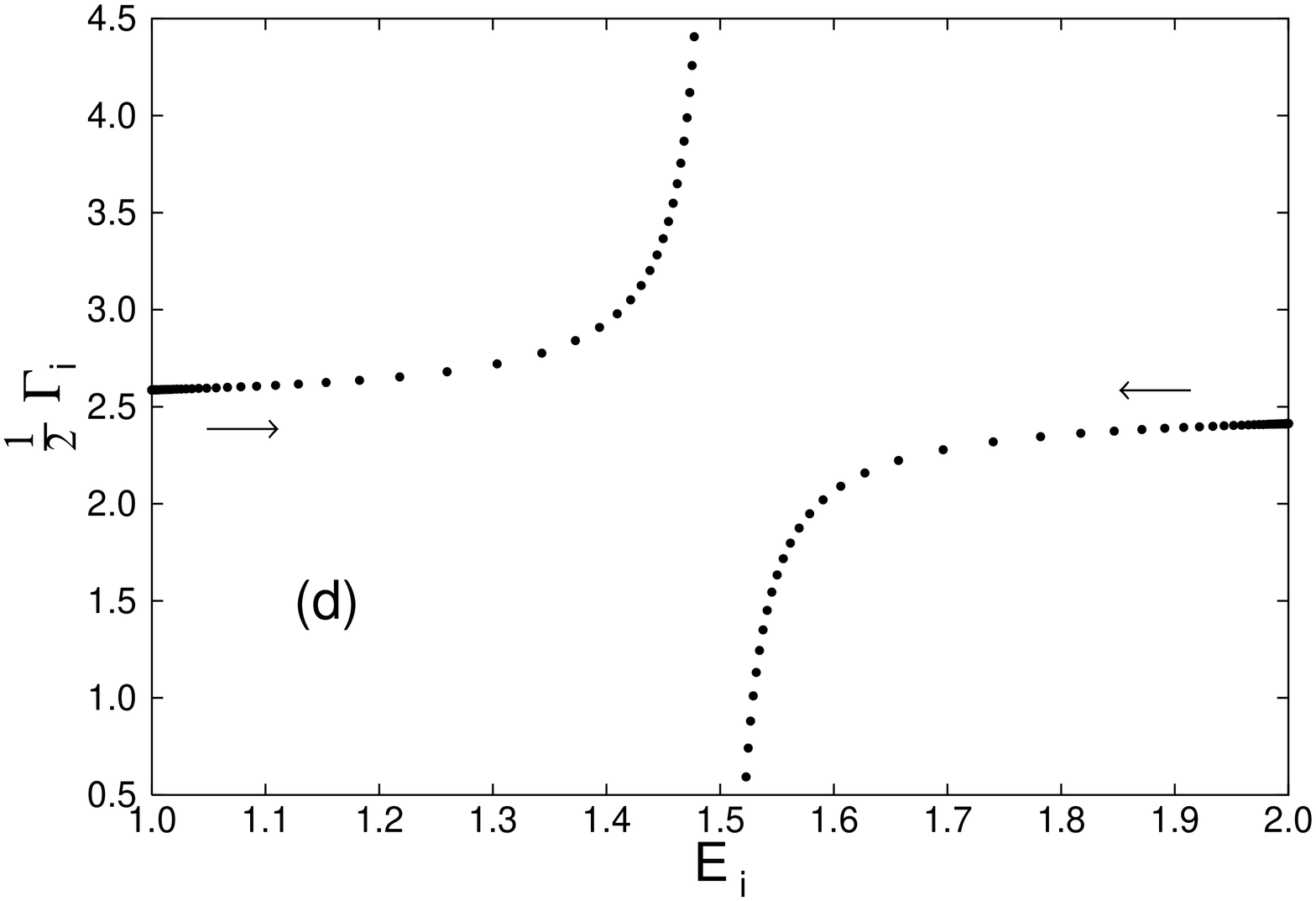,angle=-90,width=7.4cm}
\end{minipage}
\begin{minipage}[ml]{7.4cm}
\psfig{file=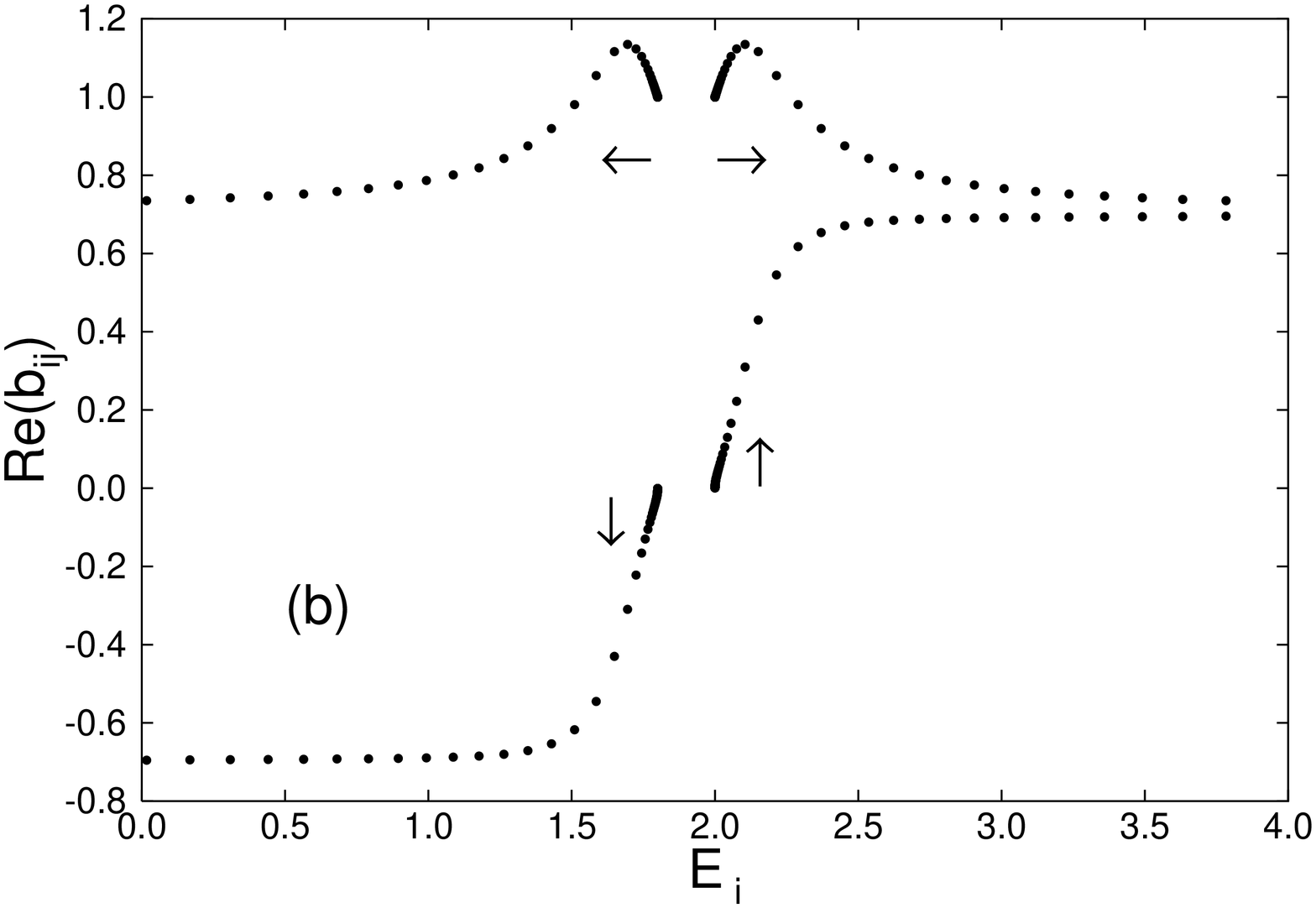,angle=-90,width=7.4cm}
\end{minipage}
\begin{minipage}[mr]{7.4cm}
\psfig{file=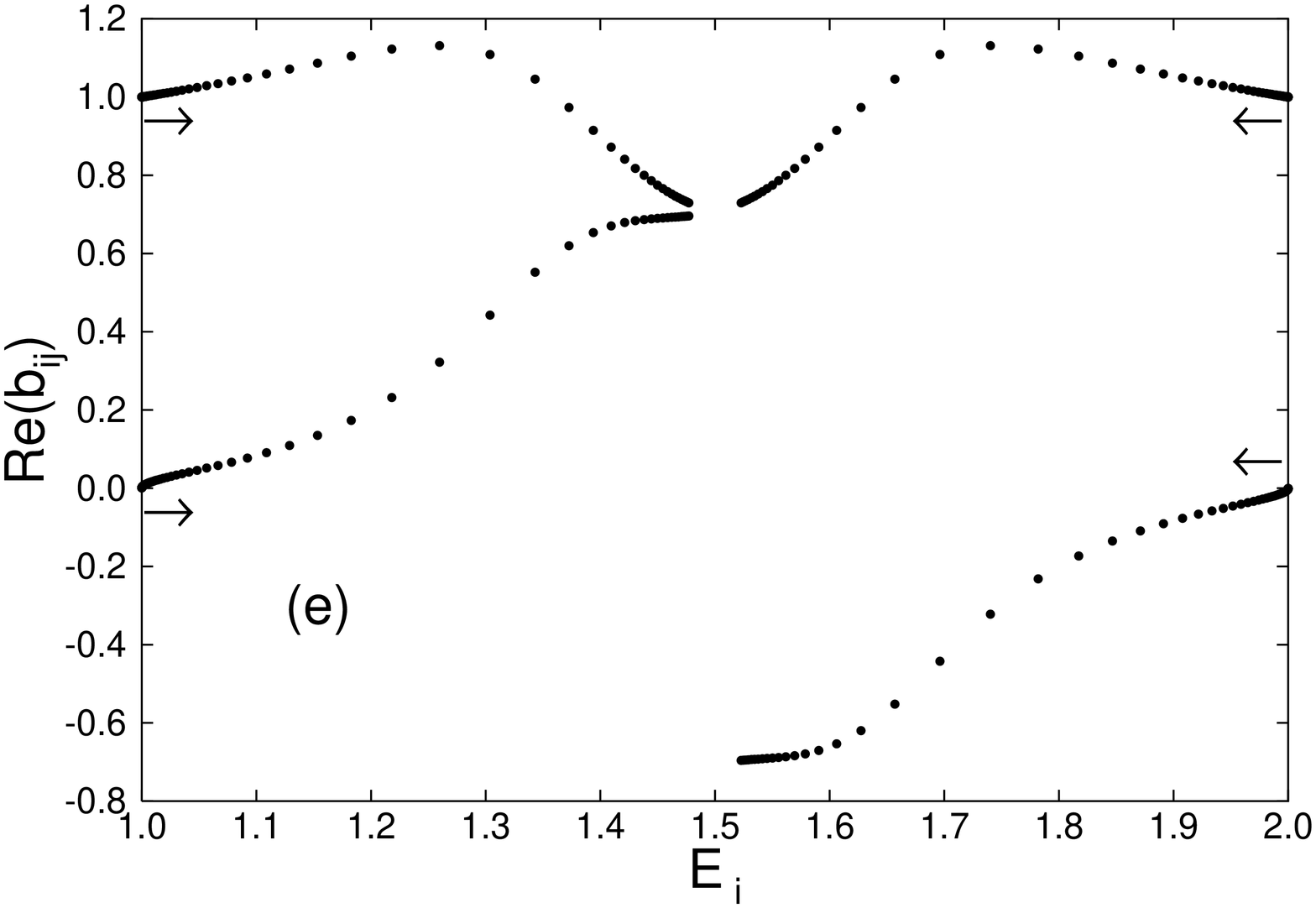,angle=-90,width=7.4cm}
\end{minipage}
\begin{minipage}[bl]{7.4cm}
\psfig{file=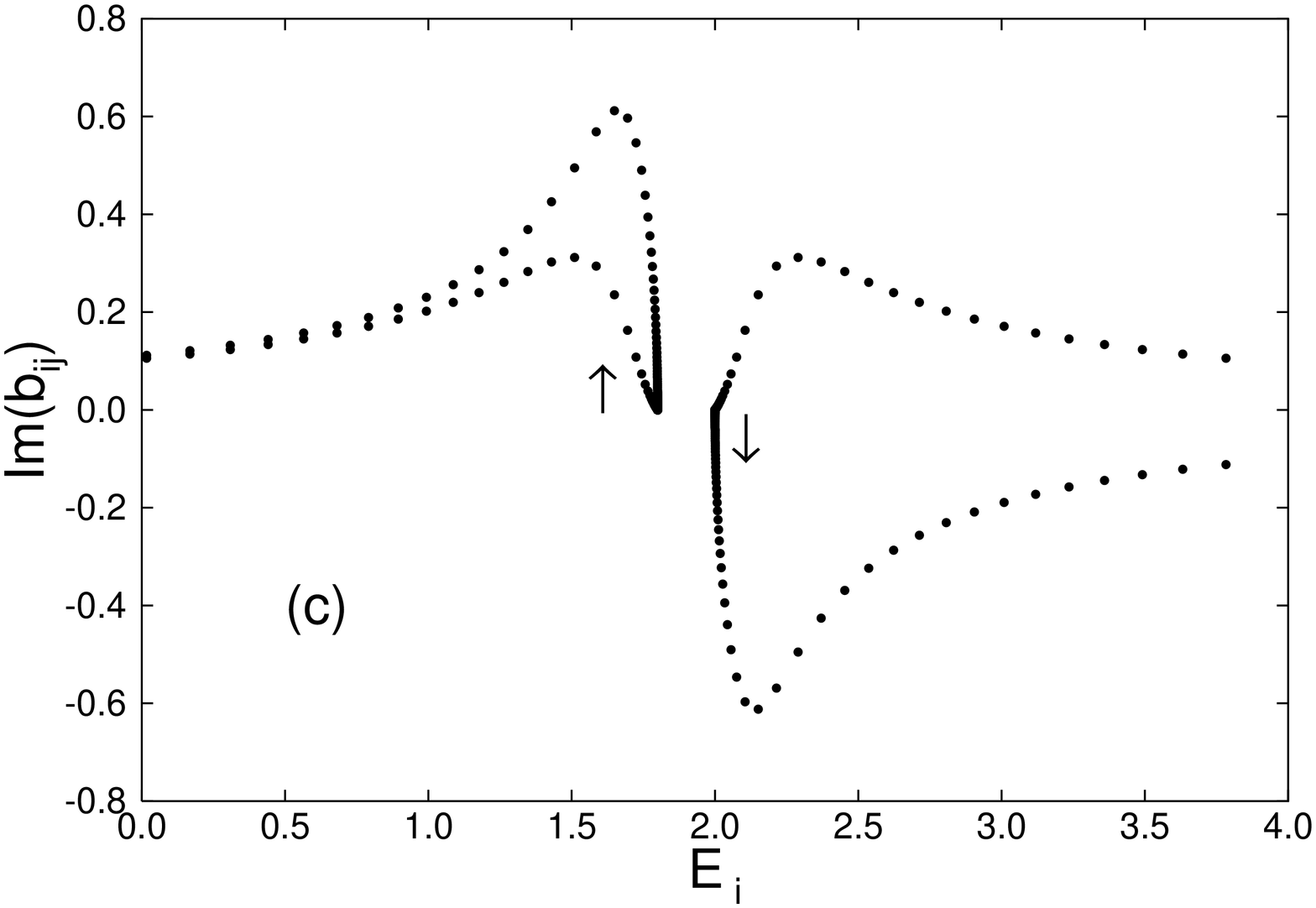,angle=-90,width=7.4cm}
\end{minipage}
\begin{minipage}[br]{7.4cm}
\hspace*{1cm}
\psfig{file=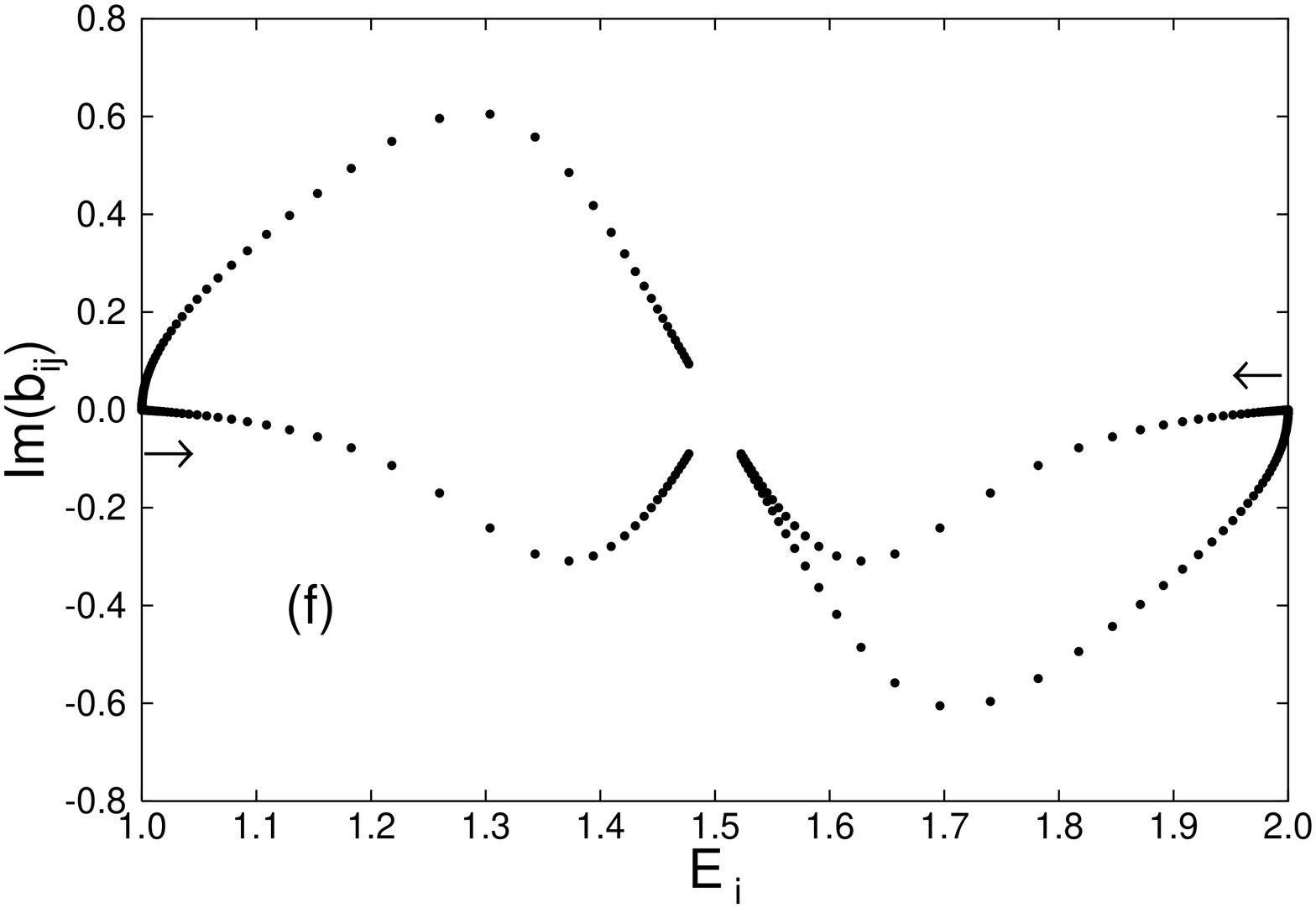,angle=-90,width=7.4cm}
\end{minipage}
\caption{{\it Eigenvalue picture:  motion of the poles of the $S$-matrix in
dependence on increasing $v_{\rm in}$ (a) and $w_{\rm ex}$ (d).
  The mixing of the wavefunctions in dependence on  
$v_{\rm in}$ (b,c)
and $w_{\rm ex}$ (e,f). At $v_{\rm in} = 0$ and $w_{\rm ex} = 0$, 
respectively, 
$b_{ij} = \delta (i,j)$ (marked by arrows).}
}
\label{fig:eigenwf}
\end{figure}

In order to illustrate the mutial 
influence of two neighboured resonance states,
we consider the following Hamilton operator
\begin{eqnarray}
{\cal H}^{(v)} & = &
\left( \begin{array}{cc} 
\epsilon_1 &  v_{\rm in}    \\
v_{\rm in} & \epsilon_2 
\end{array} \right) 
\equiv
  \left( \begin{array}{cc}
  E_{1} & v_{\rm in}  \\
  v_{\rm in}  & E_{2}
  \end{array} \right)   
 - \frac{i}{2}  \; \left( \begin{array}{cc}
 \Gamma_1  & 0 \\
  0 & \Gamma_2
  \end{array} \right)
\label{eq:avoi1}
   \end{eqnarray}               
which describes two resonance states lying at the energies $E_1$ and $E_2$.
These two states have 
the widths $\Gamma_1$ and $\Gamma_2$, respectively, and are coupled 
by   $v_{\rm in}$ (where $v_{\rm in}$ is real).
The eigenvalues are:  
\begin{equation}
{\cal E}_{\pm}^{(v)} = 
\frac{\epsilon_1 + \epsilon_2}{2} \pm \frac{1}{2} 
\; \sqrt{(\epsilon_1 - \epsilon_2)^2 + 4 v_{\rm in}^2} \; .
\label{eq:avoi2}
\end{equation}
When  $\Gamma_1 \approx \Gamma_2$, 
the coupling $v_{\rm in}$ of the two states 
leads to level repulsion along the real axis.
Numerical results for  $\Gamma_1 \not\approx \Gamma_2$ are shown in Figure 
\ref{fig:eigenwf}.a.

Let us now consider the 
Hamiltonian with the coupling $i w_{\rm ex}$ ($w_{\rm ex}$ is real) 
of the two states via the continuum,
\begin{eqnarray}
{\cal H}^{(w)} & = &
\left( \begin{array}{cc} 
\epsilon_1 &  i \; w_{\rm ex} \\
 i \; w_{\rm ex} & \epsilon_2 
\end{array} \right) 
\equiv
  \left( \begin{array}{cc}
  E_{1} & 0  \\
  0  & E_{2}
  \end{array} \right)   \;
 - \frac{i}{2}  \; \left( \begin{array}{cc}
  \Gamma_1  & - 2 w_{\rm ex} \\
 - 2 w_{\rm ex} &  \Gamma_2
  \end{array} \right) \; .
\label{eq:avoi3}
 \end{eqnarray}                  
In this case, the eigenvalues are:  
\begin{eqnarray}
{\cal E}_{\pm}^{(w)} = 
\frac{\epsilon_1 + \epsilon_2}{2} \pm \frac{1}{2}
\; \sqrt{(\epsilon_1 - \epsilon_2)^2 - 4 w_{\rm ex}^2} \; .
\label{eq:avoi4}
\end{eqnarray}
For   $E_1 \approx E_2$,
the coupling via the continuum due to $i w_{\rm ex}$  
leads to repulsion along the imaginary axis (bifurcation
of the widths), i.e. to resonance trapping.
Numerical results for  $E_1 \not\approx E_2$
 are shown in Figure \ref{fig:eigenwf}.d.

As it is well known and can be seen from eq. (\ref{eq:avoi2}), two interacting
discrete states ($v_{\rm in} \ne 0$) can not cross. In the complex plane,
however, the conditions for crossing
of two resonance states 
may be fulfilled. From  ${\cal E}_{+} = {\cal E}_{-}$, it follows
\begin{eqnarray} 
R & \equiv & (E_1 - E_2)^2 - \frac{1}{4} (\Gamma_1 - \Gamma_2)^2 + 
4 (v_{\rm in}^2 -w_{\rm ex}^2) =0
\nonumber \\
I & \equiv &(E_1 -E_2)(\Gamma_1 - \Gamma_2) + 8 v_{\rm in} w_{\rm ex} = 0
\label{eq:avoi5}
\end{eqnarray}
 for the general case of a complex
interaction $v_{\rm in} + i w_{\rm ex}$. These conditions define the
critical values of the coupling strength ($v_{\rm in}^{\rm cr}$
and $w_{\rm ex}^{\rm cr}$, respectively) at which 
the $S$-matrix has a branch point \cite{newton}.

  It is also possible that two resonance
states cross along the real or imaginary axis while the crossing is avoided
along the other axis. The conditions for such a case with $I = 0$ are 
$ R < 0 $ 
for crossing along the real axis and $R > 0$ 
for crossing along the imaginary axis. 
The crossing in the complex plane is avoided, in any case.

The trajectories for the motion of 
the eigenvalues in the complex plane
as a function of the interaction $v_{\rm in}$ and
$w_{\rm ex}$, respectively,
(Figures \ref{fig:eigenwf}.a and d) show 
the avoided crossing of the 
two resonance states in the complex plane. It occurs at a certain  critical 
value of the coupling strength. 
Here and in its neighbourhood 
a redistribution between the two states takes place. 
It is accompanied by the biorthogonality 
 of the eigenfunctions
$\tilde\Phi_i$ of  ${\cal H}$, $\; B\equiv (1/2)\sum_{i=1,2} 
\langle \tilde\Phi_i | \tilde\Phi_i \rangle >1 $.
The wavefunctions of the two resonance states
 become mixed:
$\tilde\Phi_i = \sum b_{ij} \Phi_{j}^0$ where the $\Phi_{j}^0$ are the
eigenfunctions of ${\cal H}^0$, which  is the Hamilton operator 
with 
vanishing non-diagonal matrix elements
 ($v_{\rm in} = 0$ and
 $w_{\rm ex} = 0$, respectively).
In Figures 
\ref{fig:eigenwf}.b, c and e, f,
the coefficients $b_{ij}$ are shown as a function of the coupling strength
$v_{\rm in}$ and $w_{\rm ex}$, respectively.

The results   can be summarized in the following manner.
\\[-.7cm]
\begin{itemize}
\item[--]
Weak interaction $v_{\rm in} < v_{\rm in}^{\rm cr}$ or
$w_{\rm ex} < w_{\rm ex}^{\rm cr}$:
\\
the resonance states are isolated,
their positions and widths are almost independent of the value of $v_{\rm in}$
and $ w_{\rm ex}^{\rm cr}$, respectively.\\
$B \approx 1$.
\\[-.8cm]
\item[--]
Starting at a certain critical value of the coupling strength, the behaviour
of the system is different for the two cases with 
$v_{\rm in}$ and $w_{\rm ex}$.
\\
$v_{\rm in} \approx v_{\rm in}^{\rm cr}$:
the states start to repel in energy and
to attract each other along the imaginary axis (if their 
widths at small $v_{\rm in}$ are different).\\
 $w_{\rm ex} \approx w_{\rm ex}^{\rm cr}$:
the widths bifurcate and
the levels attract each other along the real axis (if their 
positions in energy at small $w_{\rm ex}$ are different).\\
In both cases $B \gg 1$.
 The wavefunctions of the two states become
mixed. 
\\[-.8cm]
\item[--]
$v_{\rm in} > v_{\rm in}^{\rm cr}$:
the widths are almost independent
 of $v_{\rm in}$:
(equilibration  in relation to the time scale)
and the states repel each other along the real axis (level repulsion).
\\
$w_{\rm ex} > w_{\rm ex}^{\rm cr}$:
the energies are  almost independent of $w_{\rm ex}$ 
(the two states remain close to one another in their position)
and resonance trapping  occurs
(the width of one of the states   increases  with
increasing $w_{\rm ex}$, while the width of the other  state
decreases).
\\ 
In  both cases $B \to 1$. 
\end{itemize}

From a mathematical point of view, the properties of the system 
at an avoided crossing in the complex plane (i.e. in the region of the
critical coupling strength $v_{\rm in}^{\rm cr}$ and 
$w_{\rm ex}^{\rm cr}$, respectively)
 are almost the same:  repulsion of the eigenvalues along one  axis 
and attraction along the other axis. The physical meaning is, however, very
different: $v_{\rm in}$ causes  equilibrium (in relation to the
lifetime) and level repulsion along the real axis while 
$i w_{\rm ex}$ creates  different
time scales (bifurcation of the widths) 
and level attraction along the real axis.

When  the coupling contains both $v_{\rm in}$
and $i w_{\rm ex}$,
then it depends on the ratio between the two
parts whether level repulsion or attraction  along the real axis 
dominates.  In any case, the crossing of states is avoided
in the complex plane and results in a complicated interference picture. 
The wavefunctions of the resonance states are mixed permanently in the set of
the  eigenfunctions of the Hamiltonian ${\cal H}^0$
of the corresponding closed
system.    

\subsection{Realistic cases}

In \cite{pepirose}, 
the behaviour of  poles of the $S$-matrix
in an open two-dimensional regular
microwave billiard connected to a single waveguide
 is studied. As a function of the coupling
strength between the resonator  and the waveguide, the position of the
correponding resonance poles, the wavefunctions of the resonance states 
and the Wigner-Smith time delay function are calculated. 
The poles are calculated on the basis of the exterior complex scaling
method. 
The energy of the incoming wave is chosen so that only the channel 
corresponding to
the first transversal mode in the lead is open.

In \cite{pepirose}, the bifurcation of the widths (resonance trapping) can be
seen very clearly, indeed. In particular, the contraintuitive result that 
the lifetimes
of certain resonance states increase with increasing coupling to the continuum
can be traced not only in the motion of the poles of the $S$-matrix 
 in the complex plane. It can be seen also in the wavefunctions of the 
resonance states and, above all, in the measurable time-delay function. 
In the case of three interfering resonance states,  the
wavefunction of (at most) one of the long-lived trapped states may be 
pure in relation 
to the bound states of the closed resonator \cite{pepirose}. 
More exactly: $b_{ii}=1$ at 
small and large coupling strength. Some mixing of all three wavefunctions 
appears in the critical region where the wavefunctions are biorthogonal
($B>1$). 

Another example is the motion of the poles of the $S$-matrix by
varying the coupling strength between the states of an atom by means of a
laser. In \cite{marost}, the  
 positions and widths of two resonances 
in the vicinity of an autoionizing state coupled to another autoionizing one
(or a discrete state) by a strong laser field are considered.  
For different atomic parameters, 
the trajectories in the complex energy plane are traced
by fixing the field frequency $\omega$ but varying the intensity $I$
of the laser field. 
 The states are coupled directly as well as via a common continuum
and the ratio of these couplings is defined by the Fano parameter $Q$
\cite{fano}.  Most interesting is the region 
of avoided resonance crossing 
 where the motion of  each eigenvalue trajectory is  influenced strongly by
the motion of the other one.
This occurs   at a certain critical intensity $I_{\rm cr}$. 
When furthermore the frequency is equally to the critical  value 
$\omega_{\rm cr}$,
then  laser induced degenerate states arising at the double pole of the 
$S$-matrix are formed.
The strong correlation between the two states for intensities 
around  $I_{\rm cr}$ 
 reflects itself in the strong changes of the shape 
parameters of the resonances  in the cross section. 
It can therefore be traced.

In the limit of vanishing direct coupling ($Q \to 0$),
the widths bifurcate at $I=I_{\rm cr}$ as in other open quantum systems.
That means, the width of one of the resonance states increases 
with increasing $I > I_{\rm cr}$ while the  width of the other  decreases 
  relatively to the first one. In the limit $I\to \infty$, the ratio 
between the width of
the long-lived  and  short-lived state approaches zero 
 (resonance trapping). This corresponds to the situation shown in Fig. 
\ref{fig:eigenwf}.d.
In the other limiting case, the coupling via the continuum vanishes
($Q$ value large). Here, the levels repel in their energetic positions
 when $I \ge I_{\rm cr}$.
This corresponds to Fig. \ref{fig:eigenwf}.a.
 
In any case, i.e. for {\it all} $Q$ values,
the two resonance states  start to repel each other in the {\it complex} 
energy plane at $I = I_{\rm cr}$. 
The repulsion of the eigenvalues in the complex plane 
is an expression of
the strong mutual influence  of one state on the other 
one in the critical region around $ I_{\rm cr}$.  
In the transition region ($Q$ values of the order of magnitude 1),
 the trajectories show a 
complicated behaviour. Here, population trapping may appear, i.e. 
the width of one of the resonance states may vanish at a certain 
finite intensity $I_{\rm pt} > I_{\rm cr}$. It appears,  generally, 
 if the process is neither pure
 level repulsion on the real axis nor pure resonance trapping 
but the amplitudes of both processes (i.e. the direct coupling of the two
 states  and their
coupling via the continuum) are of comparable importance
and interfere.

Thus, the results obtained in \cite{marost} for two interacting atomic levels
 confirm qualitatively the results of the schematical
study represented in Section 3.1,
although not only the non-diagonal matrix elements of $H^{\rm eff}$ but also
 the width $\Gamma_1$ itself depends on $I$. These results and those for the
 microwave cavity discussed above show very clearly that individual 
resonance states can mix not only due to the two-body forces between the
 substituents of the system {\it  but also}
as a consequence of avoided resonance crossings.

Other realistic cases are considered in \cite{br}. These are the resonance 
doublet
$J^\pi = 2^+,\; T=0,\; 1$  in the nucleus $^8$Be and the $\rho -
\omega, \; T=1,\; 0$ doublet of mesons.

\section{Spectroscopic properties of an open microwave resonator}

\subsection{Calculations for the open microwave resonator}

The calculations are performed for a rectangular 
flat  resonator coupled to a waveguide.
Inside the cavity, a circular scatter is placed.
We use the Dirichlet boundary condition, $\Phi =0$, on the border
of the billiard and of the waveguide. The waveguide has a width
equal to $1$ and is attached to the resonator through a slide with
an adjustable opening
(which is described also by the Dirichlet boundary
condition). For $w=0.5$ the resonator and the waveguide are
disconnected, while $w=0$ represents the maximal coupling (opening). 

\begin{figure}
\begin{minipage}[t]{5cm}
\psfig{file=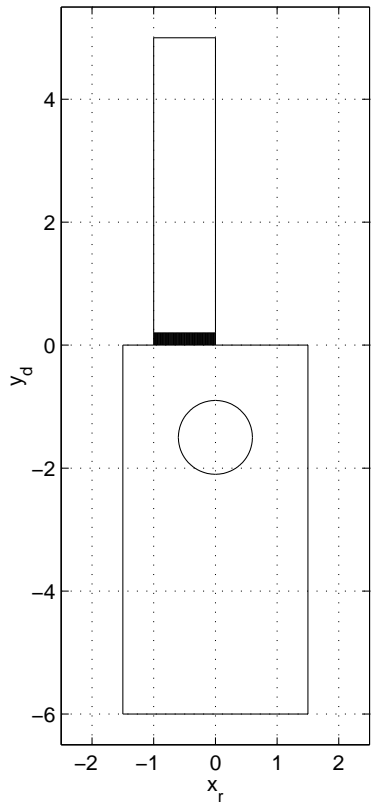,angle=0,width=5cm}
\end{minipage}
\caption{{\it The  resonator. The slide, shown in black,
 will be opened from the center to both sides (0.5 $\ge w \ge $ 0, where
$w$ = 0.5 (0) corresponds to closed (fully open)).
$x_r$ and    $y_d$ are given in arbitrary units $[x]$.    }
}
\label{fig:cavity}
\end{figure}

The cavity has a minimum area $3\times 3$ which is determined by 
$x_r= 1.5$ and $y_d=-3$ 
(compare Figure \ref{fig:cavity}). 
The area is varied by varying  $x_r$ or $y_d$ 
while both the position of the lead and the scatter inside the cavity
remain unchanged. 
 
We solve the equation
$-\Delta\Phi =E\Phi
\; .$
Inside the waveguide, the wavefunction has the asymptotic form 
$\Phi =\left(e^{iky}-R(E)e^{-iky}\right)u(x) \; .$ 
Here $u(x)$ is the transversal mode in the waveguide, $k$ is
the wave number and $R(E)$ is the reflection coefficient.
The energies and widths of the resonance states are given by the
poles of the coefficient $R(E)$ analytically continued into the
lower complex plane. They are identical to the poles of the
$S$ matrix when the fixed point equations for the $\tilde E_R$
and $\tilde\Gamma_R$
are solved (see Section 2). 

To find the poles we use the method of
exterior complex
scaling  \cite {complex,exterior}. 
For details see \cite{seromupepi}.

\subsection{Resonances as a function of the area of the resonator}

We studied the motion of the poles 
of the $S$ matrix as a function of the area
of the resonator by changing  both its length $y$ and   width $x$. 
The changes of the corresponding wavefunctions $\Phi_R$ are traced.
We studied the energy region between the two thresholds at 
 $ \pi^2 \approx 10$  and $ (2\pi)^2 \approx 40$
where only one channel is open.

\begin{figure}
\begin{minipage}[tl]{7.4cm}
\psfig{file=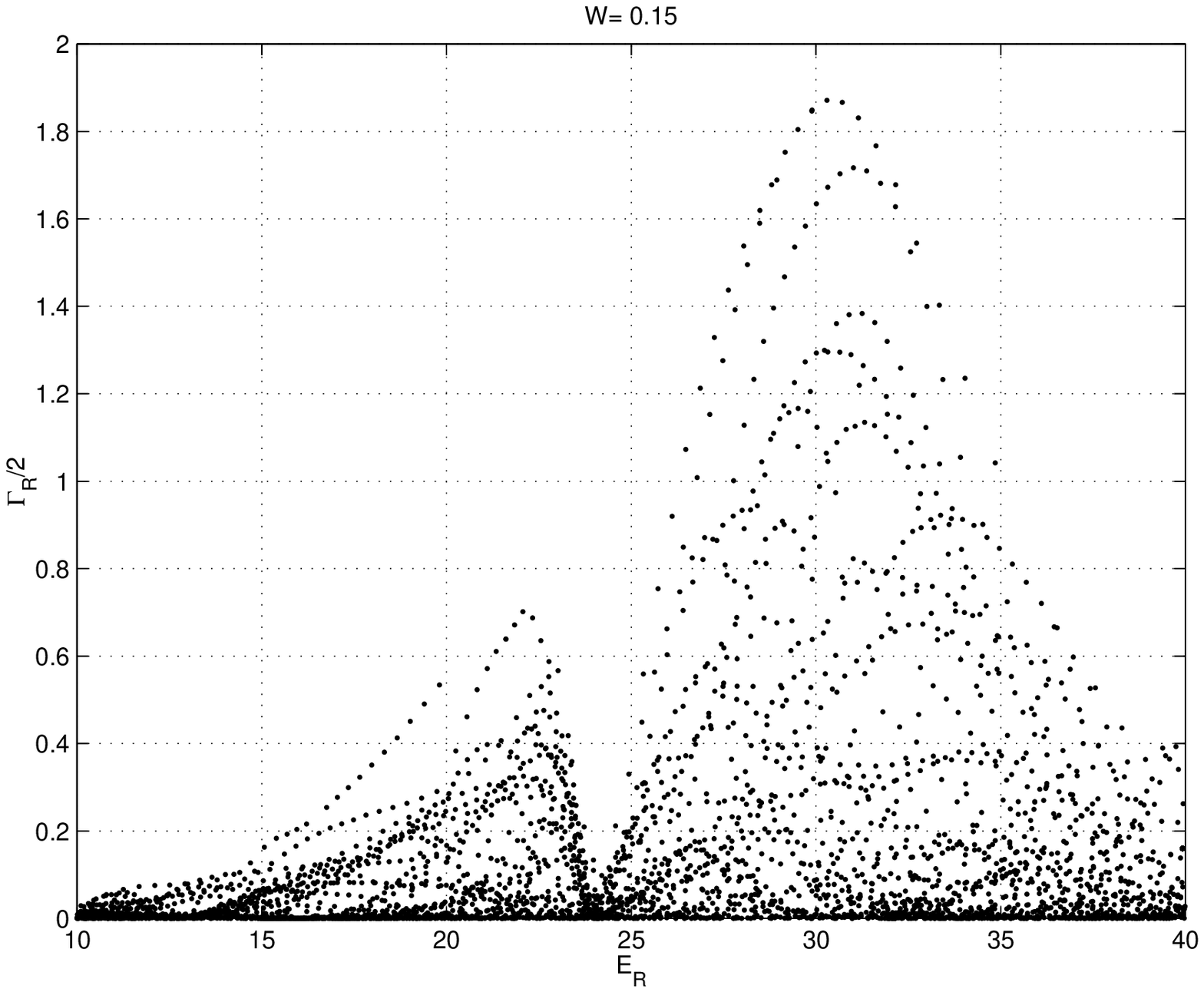,angle=0,width=7.4cm}
\end{minipage}
\begin{minipage}[tr]{7.4cm}
\psfig{file=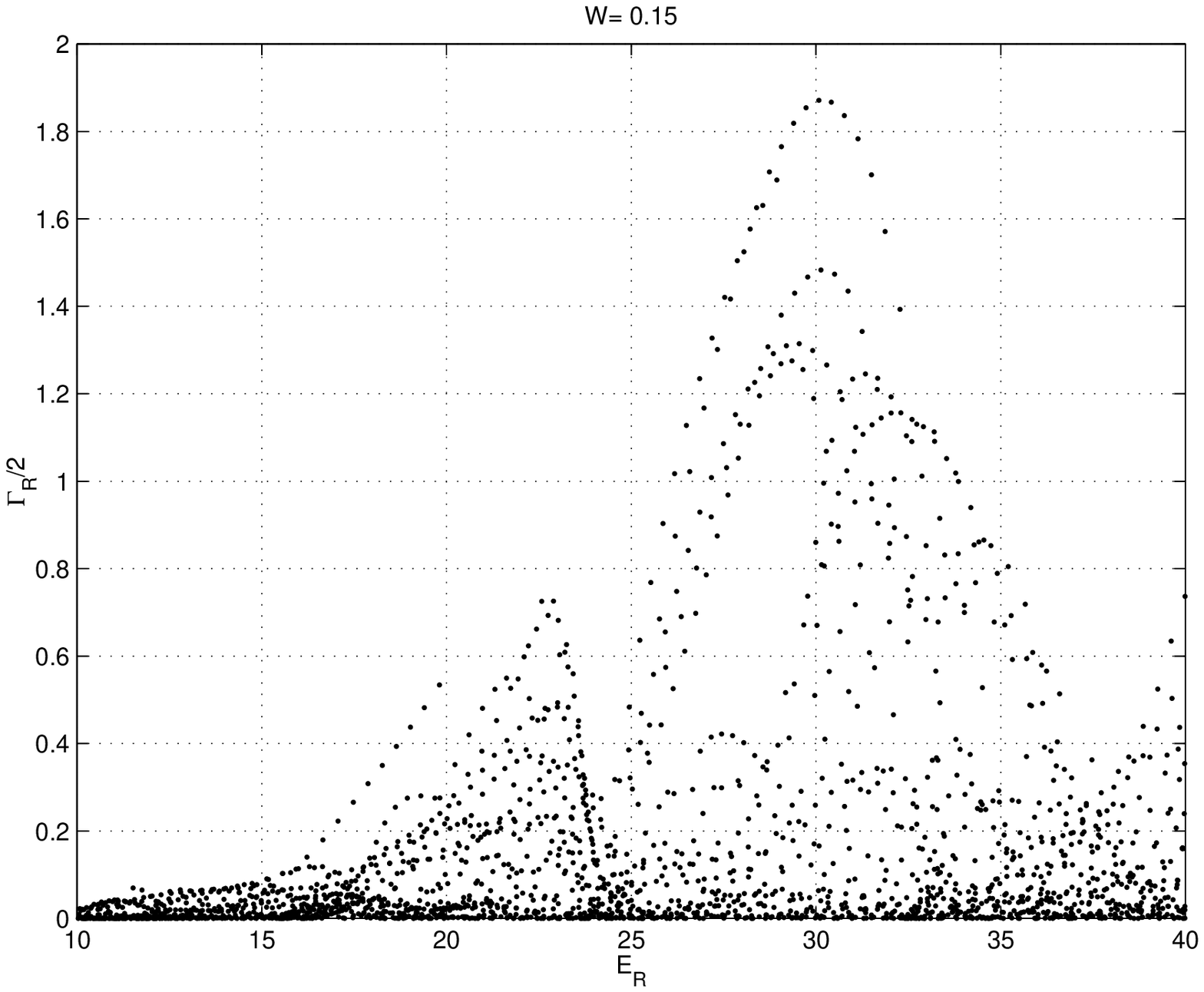,angle=0,width=7.4cm}
\end{minipage}
\begin{minipage}[bl]{7.4cm}
\psfig{file=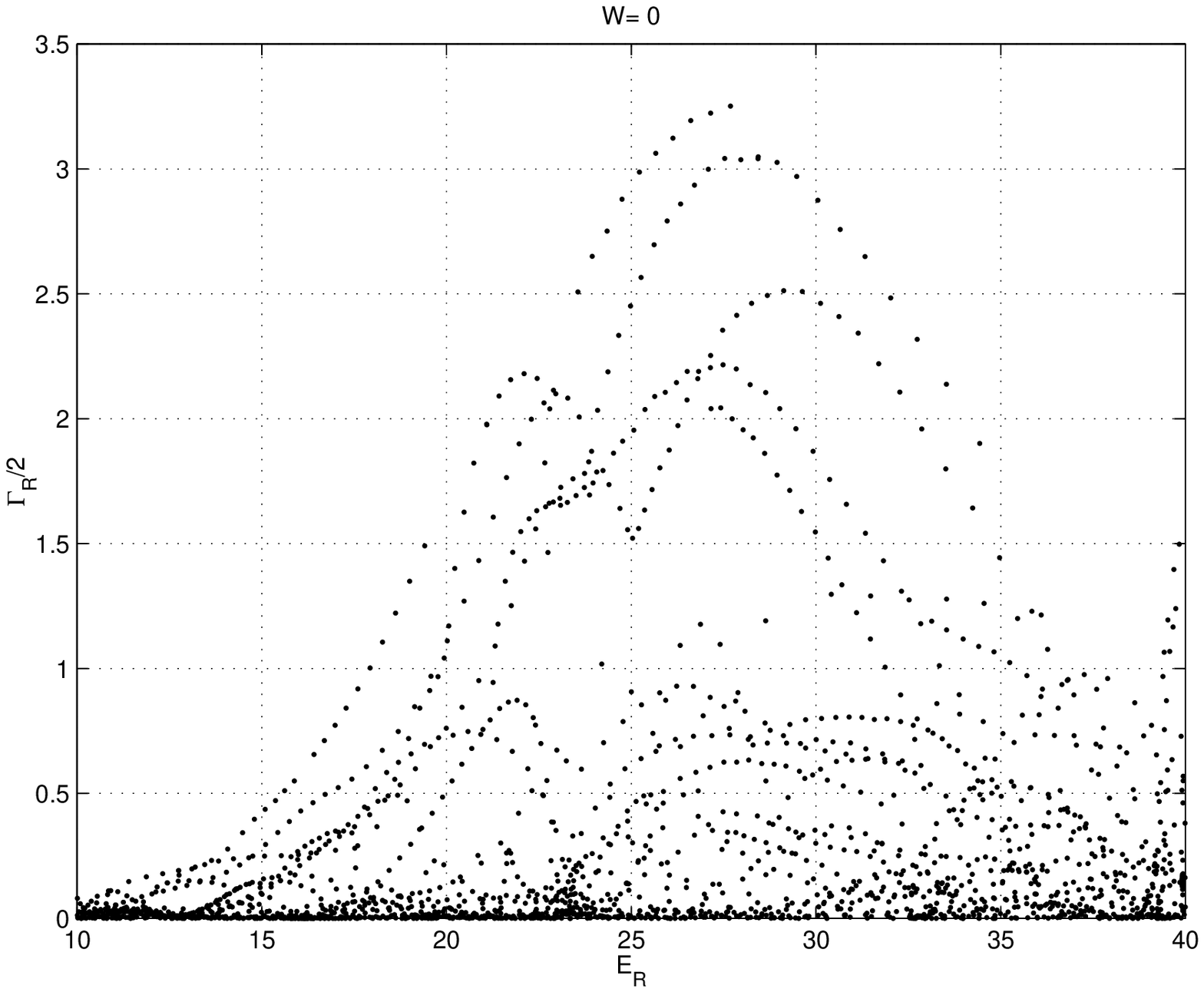,angle=0,width=7.4cm}
\end{minipage}
\begin{minipage}[br]{7.4cm}
\hspace*{1cm}
\psfig{file=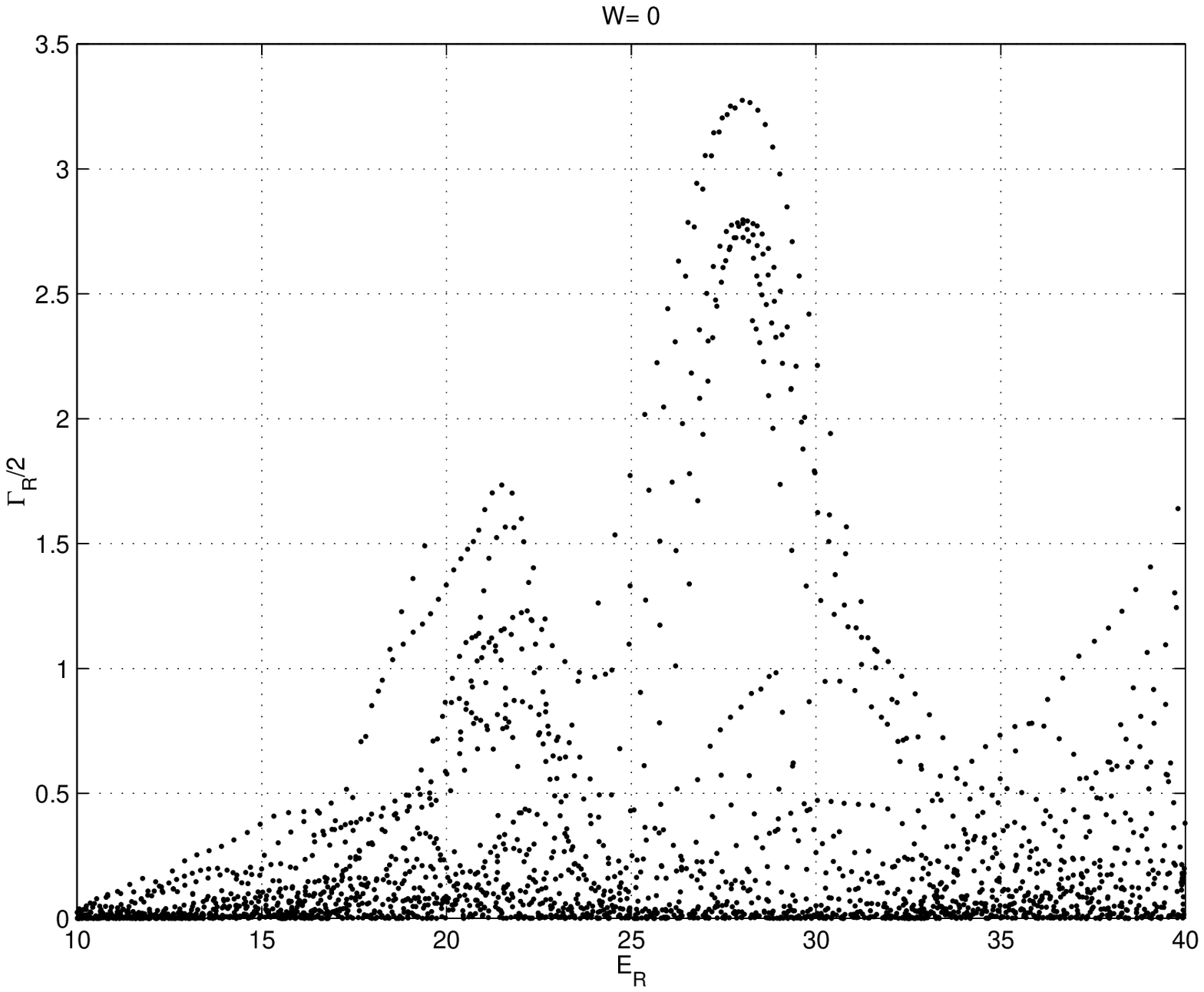,angle=0,width=7.4cm}
\end{minipage}
\caption{{\it Eigenvalue picture:  motion of the poles of the $S$-matrix in
dependence on increasing length (left) ($y_d$ = -6.0 : 0.02 : -3.0; 
 $x_r$ = 1.5)
and width (right)   ($x_r$ = 1.5 : 0.02 : 3.5;   $y_d$ = -3.0)
of the resonator.
The opening of the aperture is $w$ = 0.15 (top) and $w$ = 0 (bottom).
The energies are given in units of $[x]^{-2}$.}
}
\label{fig:eigenxy}
\end{figure}

In Figure \ref{fig:eigenxy}, the eigenvalue picture 
is shown for 
$w=0.15$ (aperture partly closed by the slide) 
and $w=0$ (aperture fully open) 
for different values of the
length and width of the resonator.
In all cases, oscillations of the widths as a function of $y_d$ or $x_r$ 
in the energy region considered 
can be seen. The amplitudes of the oscillations are  larger for larger widths. 
At $w=0.15$, all  states corresponding to different $y_d $  ~($-3 > y_d > -6$) 
and lying around 24  have small widths. This  
is caused, obviously,  by 
certain symmetry properties of the wavefunctions
in relation to the channel (since this energy is in the middle between the two
thresholds). 
The minimum in the widths vanishes when the wavefunctions are strongly  mixed
via the continuum ($w=0$). This shows that the coupling to the channel 
washes out some spectroscopic  properties of the closed system.

For $w=0.15$, the energies and widths of the states lying in the energy 
region around 24   are shown in Figure   \ref{fig:eigen10y}
 as a function of $y_d$. 
The $E_R(y_d)$ show typical avoided crossings while the picture of the
$\Gamma_R(y_d)$ is more complicated. For the three states denoted by 
diamonds, stars and circles, respectively, 
the wavefunctions are shown in Figure  \ref{fig:wf10y3} for 
10 different neighboured
 values of $y_d$. 

\begin{figure}
\begin{minipage}[t]{15cm}
\psfig{file=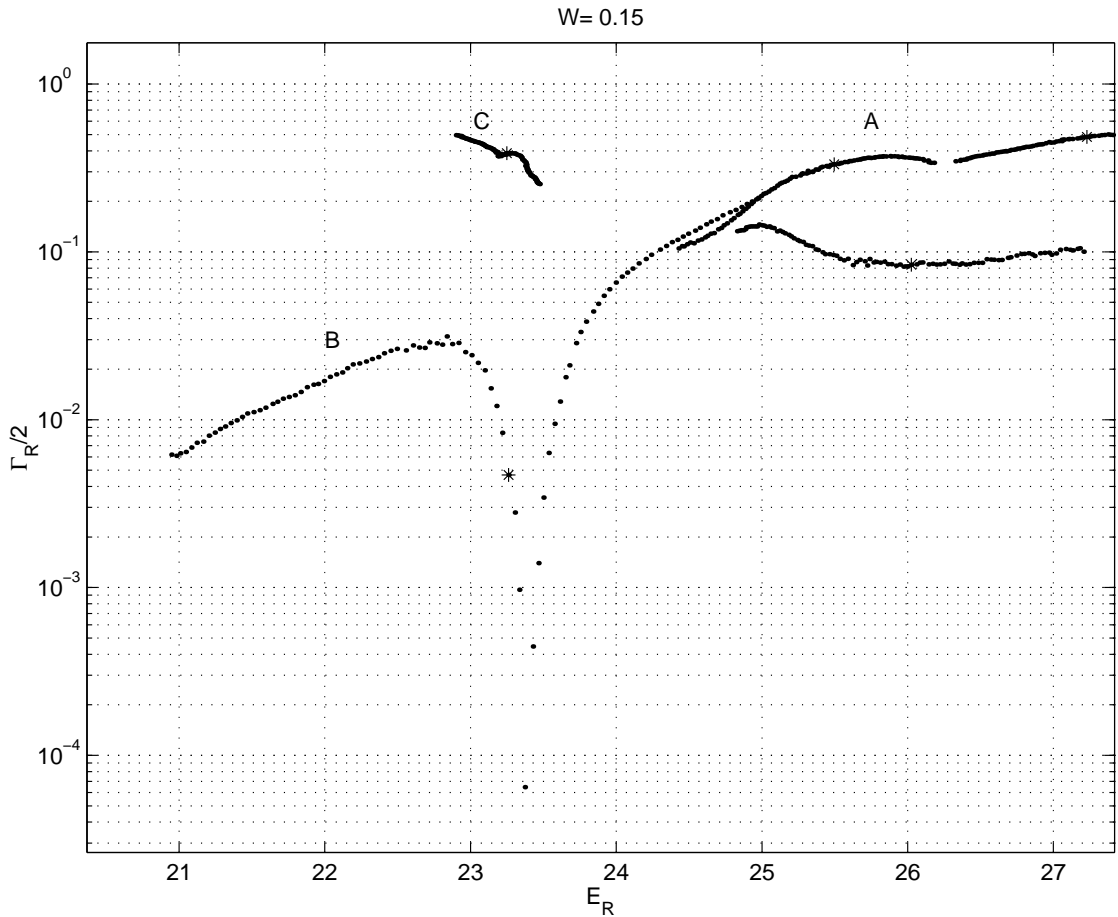,angle=0,width=15cm}
\end{minipage}
\caption{{\it 
 Eigenvalue picture:  motion of the poles of the $S$-matrix in
dependence on increasing length of the resonator  ($y_d$ =
-3.75 : 0.001 : -3.65). 
 The eigenvalues at $y_d$ =  -3.693 are marked by stars.}
}
\label{fig:free}
\end{figure}

 Two states ($B$ and $C$) 
cross  freely in the energy  at $E_R \approx 23$.
The wavefunctions of the two states $B$ and $C$  
are  very different from one another and the 
interaction due to $\Re(W)$ between them is obviously small.
The wavefunctions of both states do almost not change
in the crossing region. Only in the widths, some repulsion can be seen
 caused obviously by $\Im(W)$. This can be seen from
Figure \ref{fig:free} where the results are shown from a  calculation 
with smaller 
steps in $y_d$ around the free crossing.

Around  $E_R=27$, the
state $B$ avoids crossing in energy with the other state ($A$) 
 at some value $y_d^{\rm cr}$ (around -3.63). In this
region, the wave functions of both states become strongly mixed,  their widths
become comparable and cross.  The repulsion in their energies can be seen.
The avoided crossing is caused mainly by $\Re(W)$.
Beyond the critical region, the wavefunctions of the two states  remain  mixed
although some hint to their exchange can be seen.

These results show that avoided level crossing in the complex plane can 
be seen 
in the projection onto the energy axis or in the projection onto the width
axis.
In the first case, $\Re(W)$ dominates while in the 
second case, the mixing
of the states occurs mainly due to $\Im(W)$.

\begin{figure}
\begin{minipage}[t]{17cm}
\psfig{file=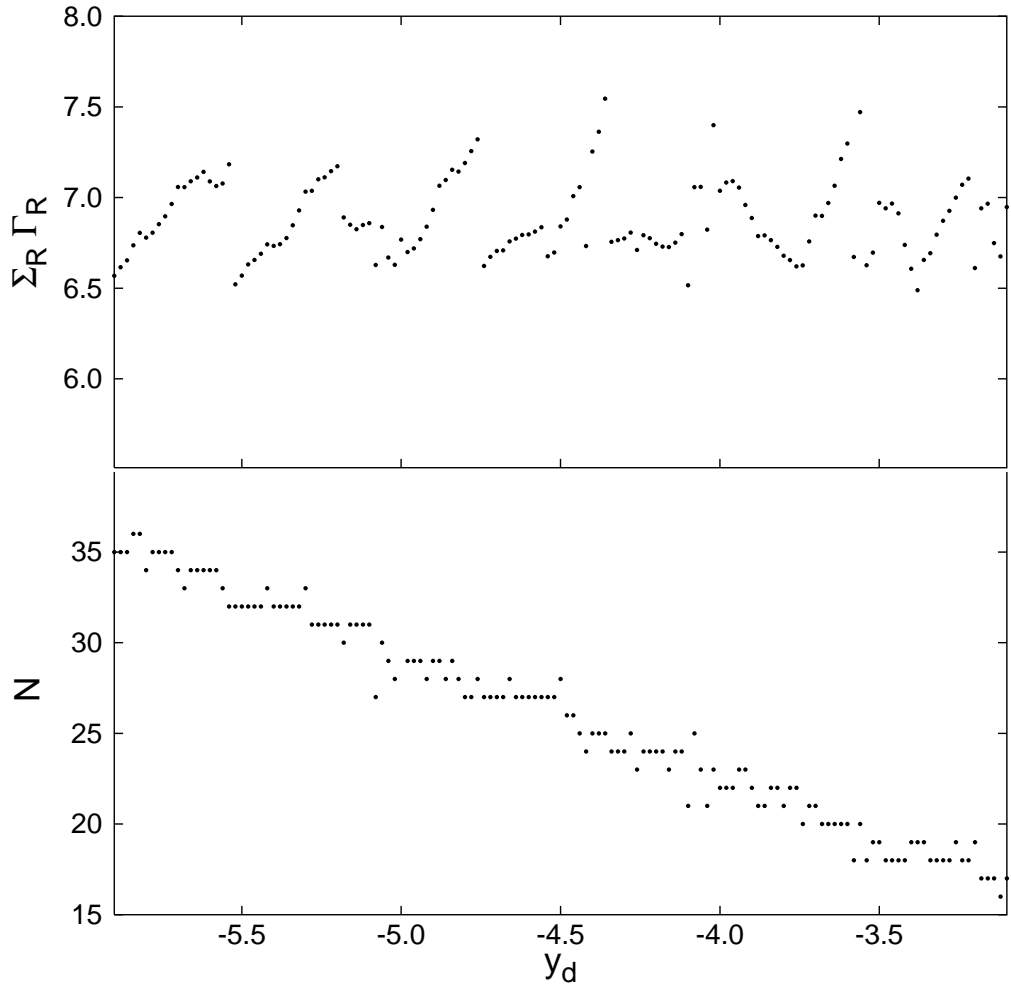,angle=-0,width=17cm,bbllx=50pt,bblly=330pt,bburx=556pt,bbury=750pt,clip=}
\end{minipage}
\caption{{\it 
The sum $\sum_R \Gamma_R$ of the widths (top) and the number $N$ (bottom)
  of the states lying between 
the two thresholds  shown  as a function of 
the length $y_d$ of the cavity. $w=0.15$.}
}
\label{fig:gasum}
\end{figure}

According to the oscillations of the widths
and the varying number of states as a function of the length or
width of the resonator, the sum of the widths of all states, 
lying between the two thresholds,  
fluctuates as a function of these values.
In Figure \ref{fig:gasum} (bottom), we show 
the number $N$ of states as a function of $y_d$ (for $w= 0.15$).
This number increases since the number of states moving from above into the 
energy region
considered is larger  than the number of states leaving it
to get bound.
On the average,  $\sum_R \Gamma_R$ is constant for a fixed value of $w$
with fluctuations smaller than  10 \%.
This can be seen from the
example with $w= 0.15$  shown in Figure \ref{fig:gasum} (top).
The coupling of the cavity to the lead is therefore
characterized by $w$ but not by the area of the cavity.

\subsection{Resonances as a function of the coupling strength to the lead}

\begin{figure}
\begin{minipage}[t]{12.5cm}
\psfig{file=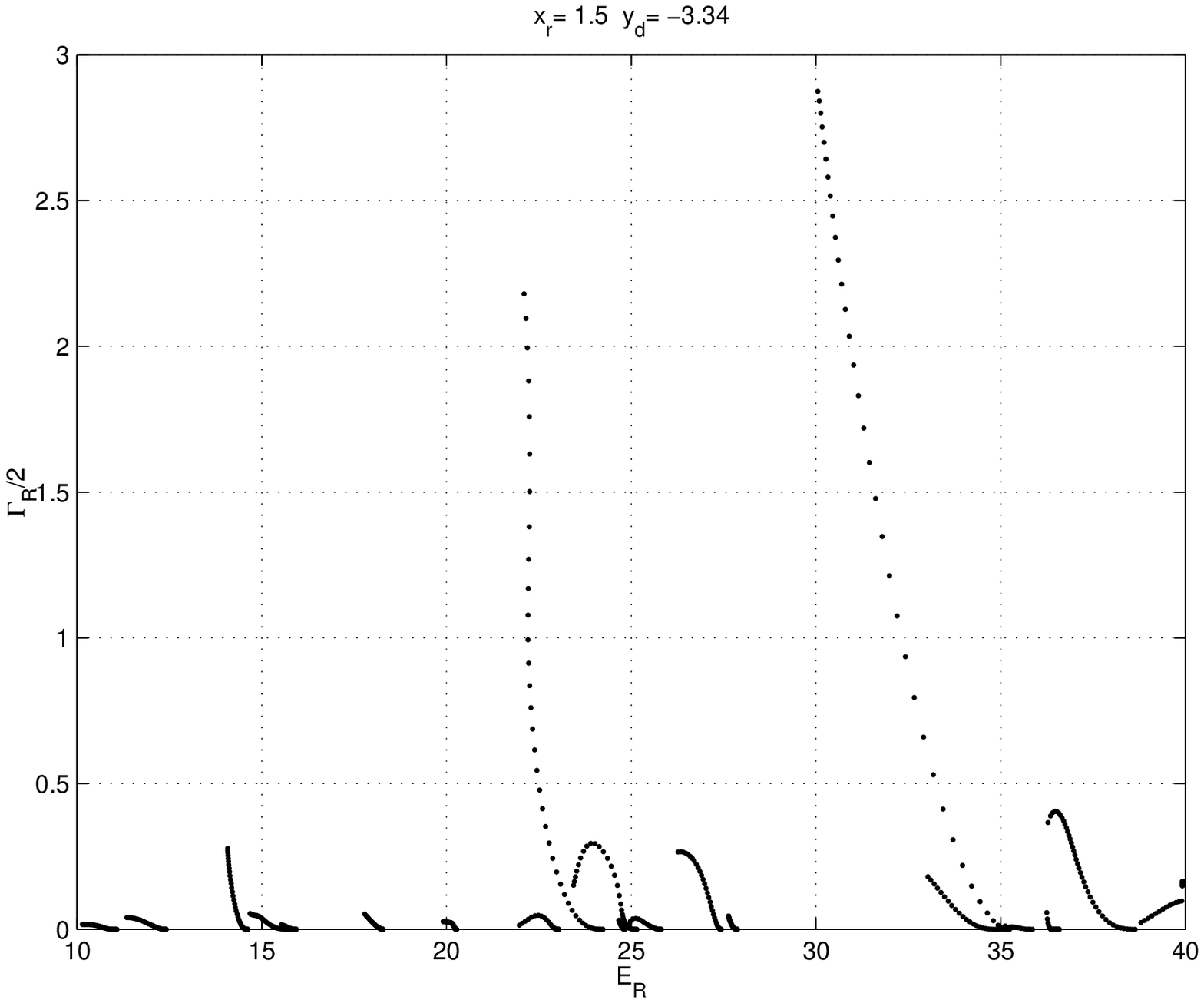,angle=0,width=12.5cm}
\end{minipage}
\begin{minipage}[b]{12.5cm}
\psfig{file=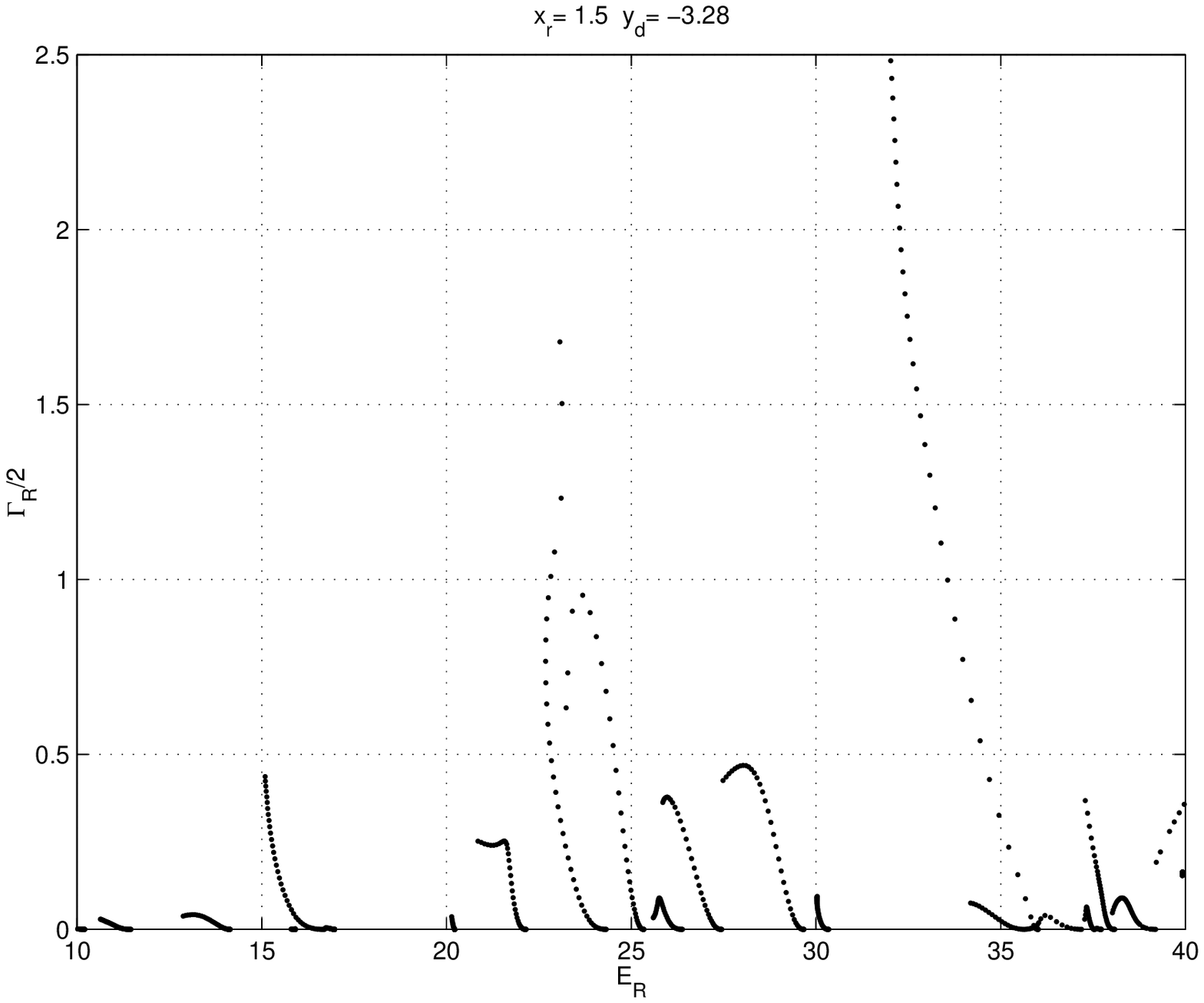,angle=0,width=12.5cm}
\end{minipage}
\caption{{\it Eigenvalue picture:  motion of the poles of the $S$-matrix in
dependence on increasing opening
(decreasing $w, 
~w = 0.4 : 0.01 : 0)$
for $x_r$ = 1.5, $y_d$ = -3.34 (top) and  $y_d$ = -3.28 (bottom). }
}
\label{fig:eigenw}
\end{figure}

In Figure  \ref{fig:eigenw}, we show the eigenvalue picture 
obtained by varying 
$w$ from 0.4 (almost closed aperture) to 0 (fully open aperture).
The width of the resonator is determined by $x_r=1.5$ and the
length  by the two neighboured values
$y_d=-3.34$ (Figure \ref{fig:eigenw} top) and 
 $y_d=-3.28$ (Figure \ref{fig:eigenw} bottom).
In both cases, collective states are formed. They are formed in regions where 
the level density is comparably high. Even at full opening of the aperture,
the collective states belonging to the different groups
do not overlap. Thus, they scarcely mix  via the
continuum. 

In Figure \ref{fig:wf3y}, we show the wavefunctions of the 
collective states 
from the lower part of Figure  \ref{fig:eigenw}.
Although the wavefunctions of the collective  states are
very different from one another at small opening of the aperture $(w=0.4)$, 
they are similar at full opening $(w=0)$ (Figures  \ref{fig:wf3y}.d,f)
where they have large amplitudes  near to the aperture.
The state shown in the middle (Figure   \ref{fig:wf3y}.e)
is trapped by the state to the left (Figure  \ref{fig:wf3y}.d) 
at a comparably large 
opening (compare Figure  \ref{fig:eigenw} bottom).  
The  wavefunctions of the collective states at $w=0$ in the long
and in the broad resonators ($x_r \to 4.0, \; y_d \to -6.0$) are 
also similar to 
those shown in Figures  \ref{fig:wf3y}.d and  \ref{fig:wf3y}.f.

\subsection{Resonances and conductance of the resonator} 

The conductance of the resonator is described by the matrix elements
$S_{cc'}$, equation (\ref{eq:smares}), where $c$ is the channel of the
 incoming wave and
$c'$ that of the outgoing wave. 
In our calculations, the second lead is on the lower right corner of the
cavity, symmetrically to the first lead on the upper left corner. It is
$x_r=1.5$ and $y_d=-3$. 
 
\begin{figure}
\begin{minipage}[tl]{7.4cm}
\psfig{file=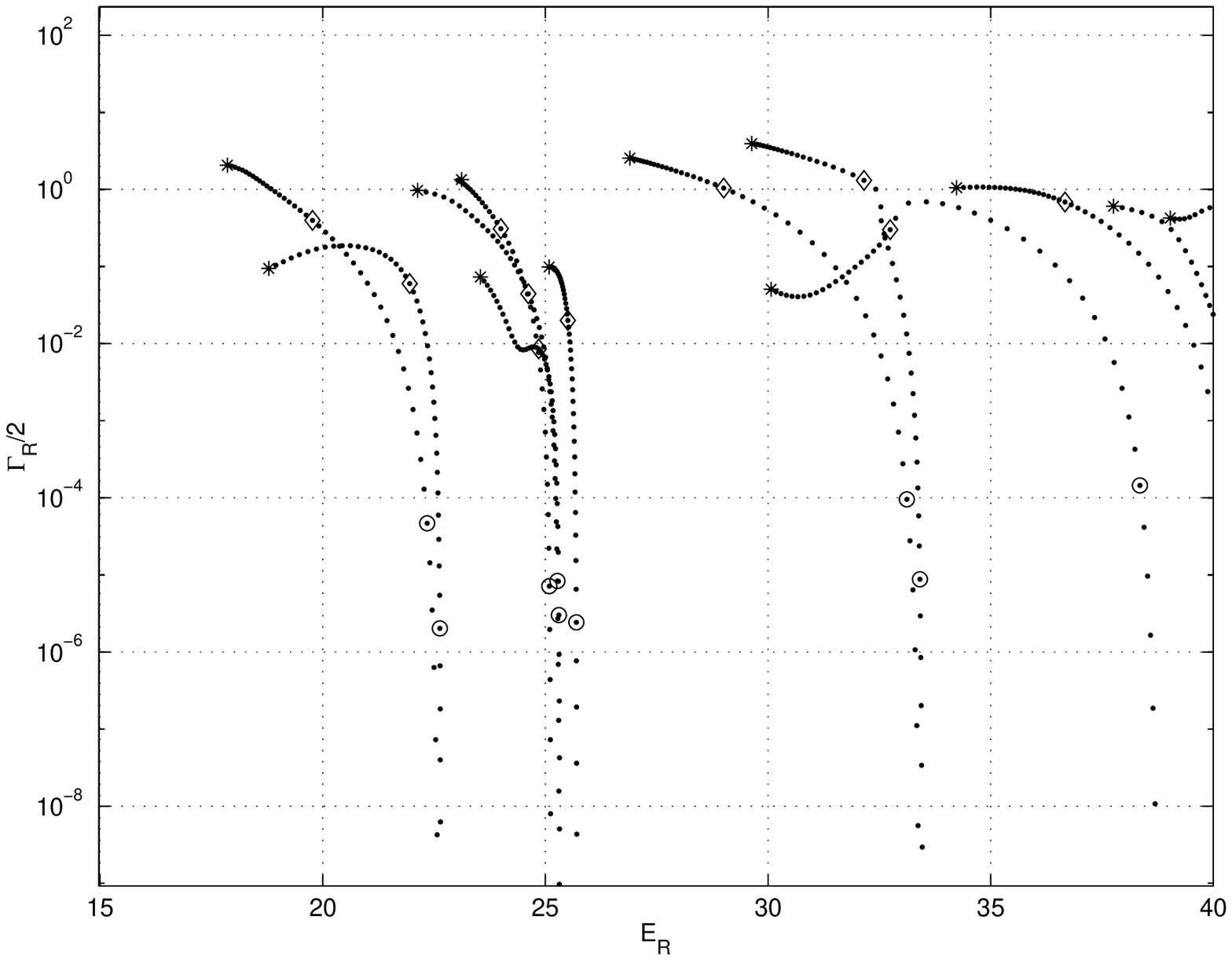,angle=0,width=7.4cm}
\end{minipage}
\begin{minipage}[tr]{7.4cm}
\psfig{file=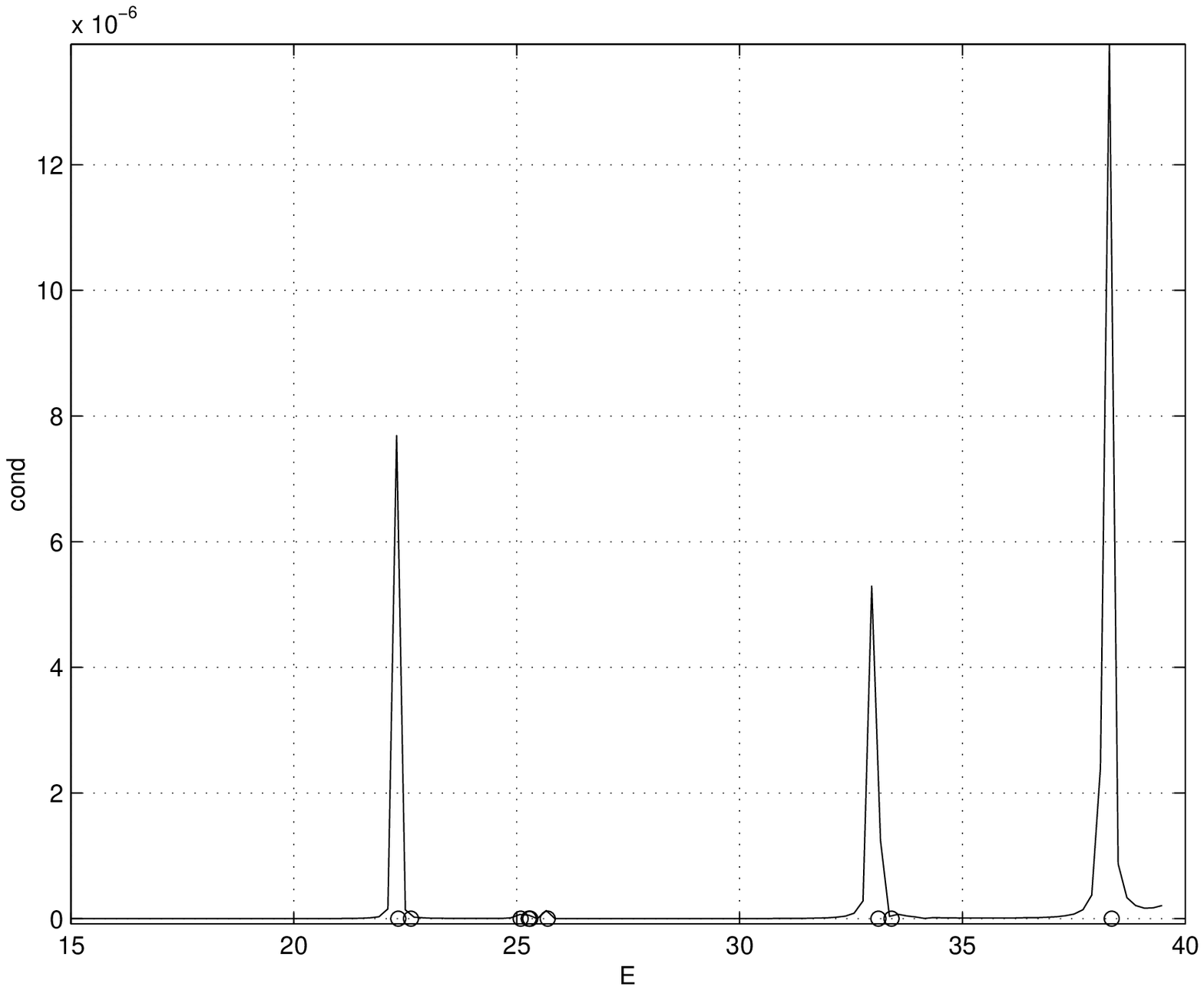,angle=0,width=7.4cm}
\end{minipage}
\begin{minipage}[bl]{7.4cm}
\psfig{file=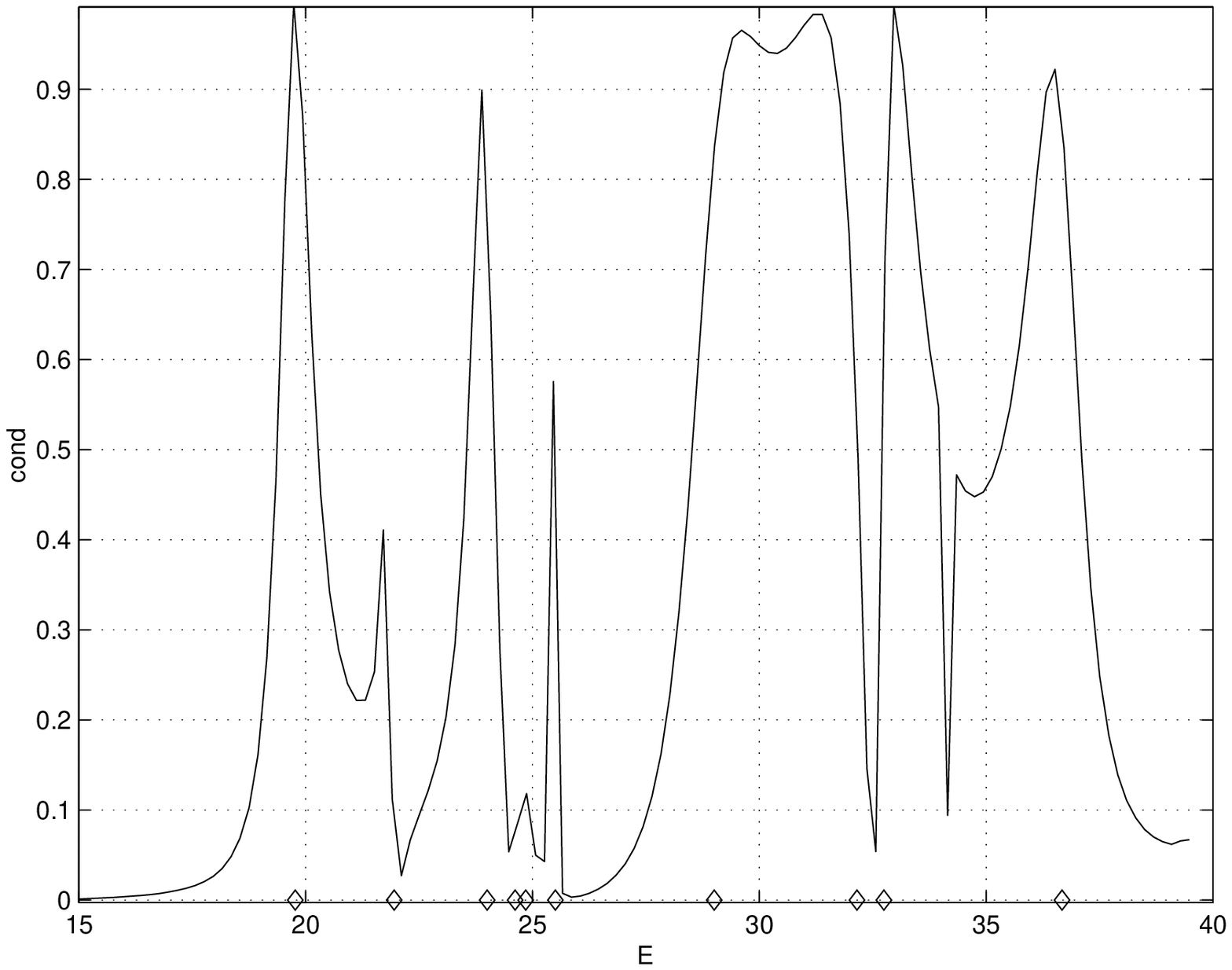,angle=0,width=7.4cm}
\end{minipage}
\begin{minipage}[br]{7.4cm}
\hspace*{1cm}
\psfig{file=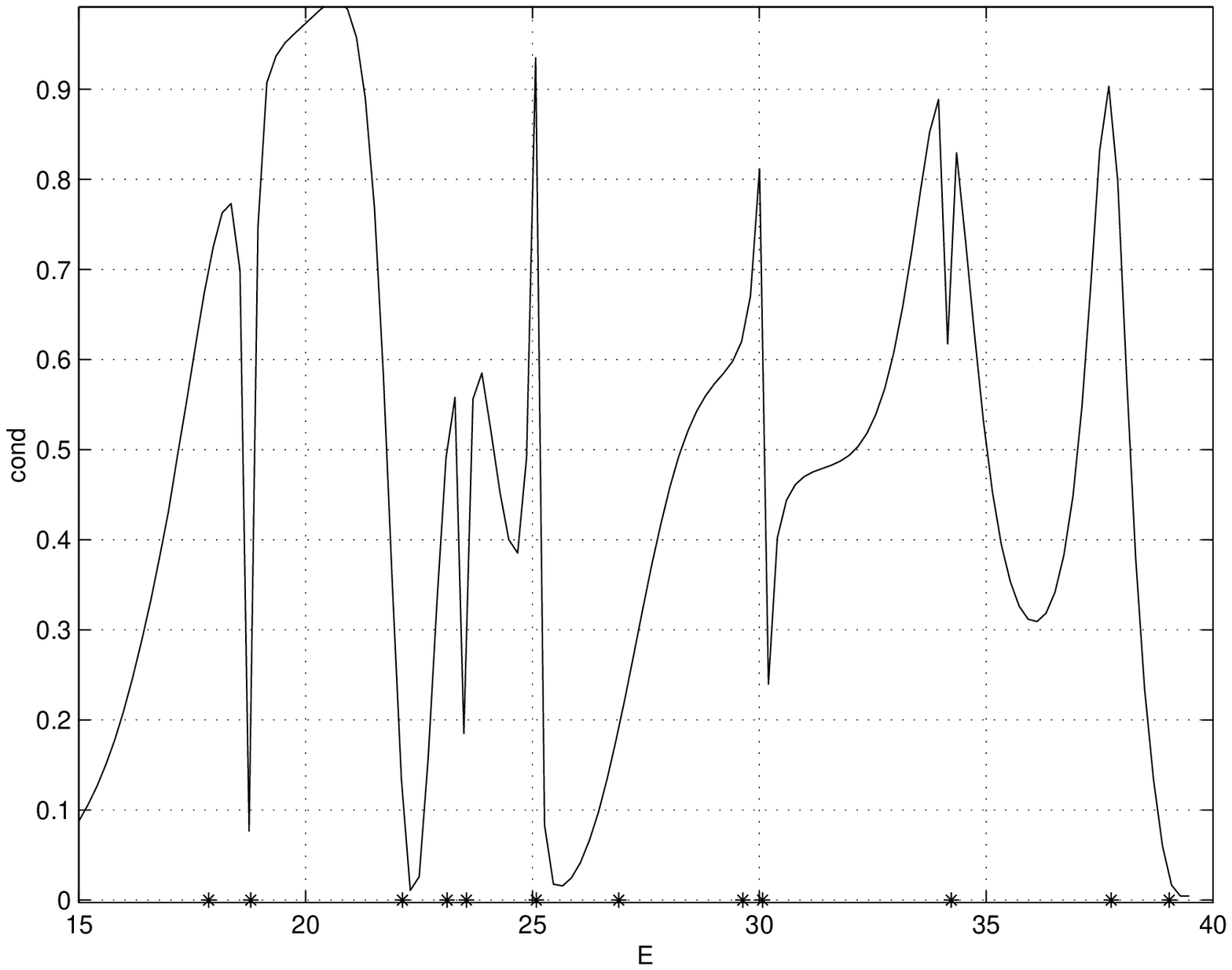,angle=0,width=7.4cm}
\end{minipage}
\caption{{\it Eigenvalue picture:  motion of the poles of the $S$-matrix in
dependence on increasing opening
(decreasing $w$, top left) and the conductance as a function of
$E$ for $w=$ 0.4 (top right), ~0.2 (bottom left) and 0 (bottom right). 
 The values $E_R - (i/2)
\Gamma_R$ and $E_R$, respectively, for  $w=$ 0.4,
~0.2 and 0 are marked by circles, diamonds and stars.
}
}
\label{fig:condw}
\end{figure}

In Figure \ref{fig:condw}, the conductance at three different coupling
strengths to the leads
is shown together with the eigenvalue picture.
In the eigenvalue picture, one can see the formation of two short-lived 
states at large opening (small $w$) in
each group. This corresponds to the coupling of the resonator to two leads.   
It can be seen from the wavefunctions of the  states 
that {\it both} short-lived states of each group are coupled
strongly to {\it both} leads. The conductance is therefore large at 
large opening.

 At low opening
($w=0.4$), the conductance peaks coincide with the resonance peaks. At larger
opening ($w=0.2$ and 0), the conductance is an interference picture created by
 the overlapping resonances. The influence of the short-lived
resonances onto the conductance can clearly be seen.  
 
\begin{figure}
\begin{minipage}[t]{15cm}
\psfig{file=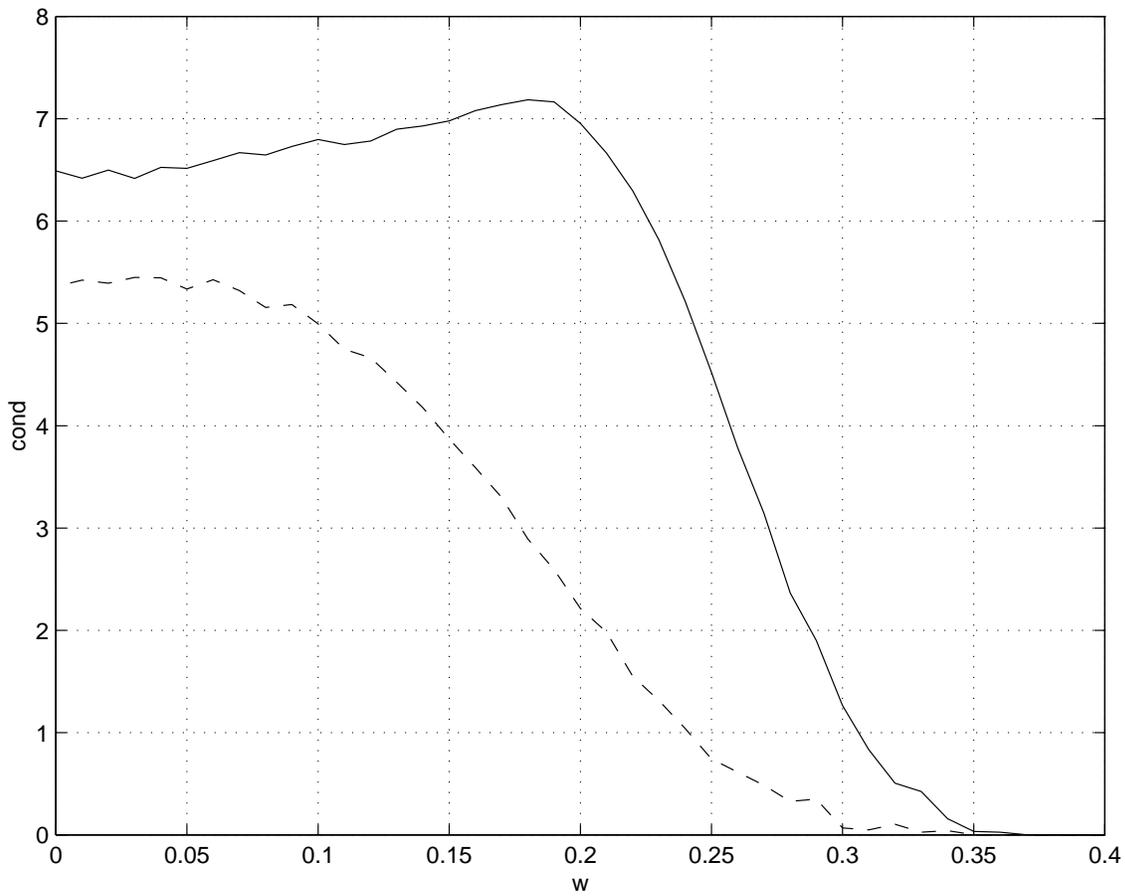,angle=0,width=15cm}
\end{minipage}
\caption{{\it 
Integrated conductance as a function of $w$ in the energy region  
25 $\le$ E $\le$ 40
(full line) and  15 $\le$ E $\le$ 25
(dashed line). 
}
}
\label{fig:condint}
\end{figure}

In Figure \ref{fig:condint}, the conductance is  integrated over the energy
of each group ($15 \le E \le 25$ and $25 \le E \le 40$) and 
plotted as a function of $w$. The conductance at the higher energy 
increases quite rapidly in a small region of $w$ which corresponds to the
critical region around $w_{\rm cr}$
(compare \cite{seromupepi}).

\section{Discussion of the results}

As demonstrated in Sections 3 and 4, 
the wavefunctions of a
quantum system mix under the influence of 
the hermitian as well as the anti-hermitian part of the Hamiltonian.
If the hermitian part of the Hamiltonian is dominant then
avoided crossing can be seen along the energy axis ({\it level 
repulsion}).  
If the anti-hermitian part of the
coupling via the continuum becomes important, then  
resonance trapping ({\it bifurcation of the widths}) appears.
In general, both types of mixing 
appear and may interfere.
Note that this interaction between different states of a quantum system
via the continuum
does not require two-body forces between the constituents
of the system. 

\begin{figure}
\begin{minipage}[t]{12.5cm}
\psfig{file=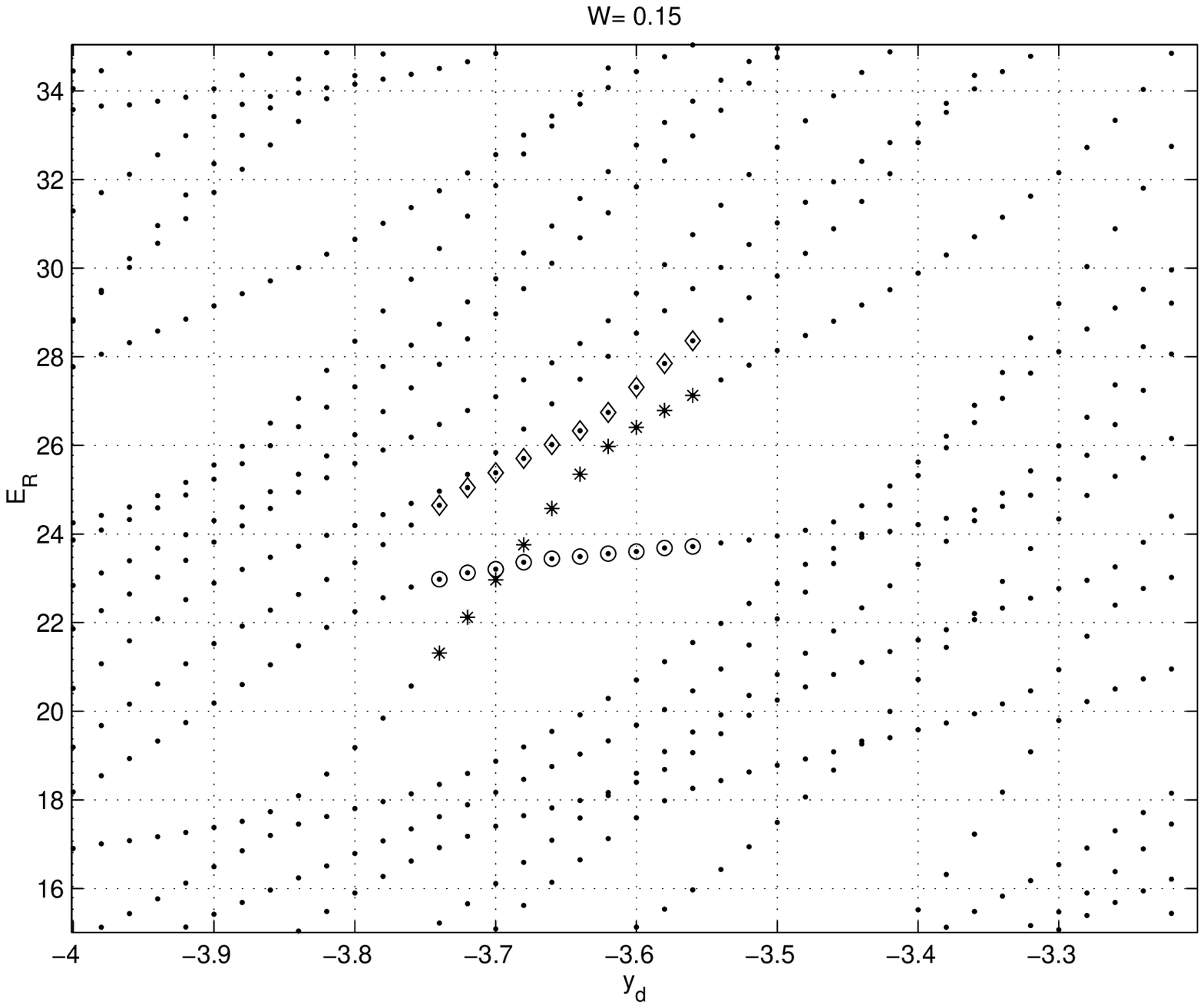,angle=0,width=12.5cm}
\end{minipage}
\begin{minipage}[b]{12.5cm}
\psfig{file=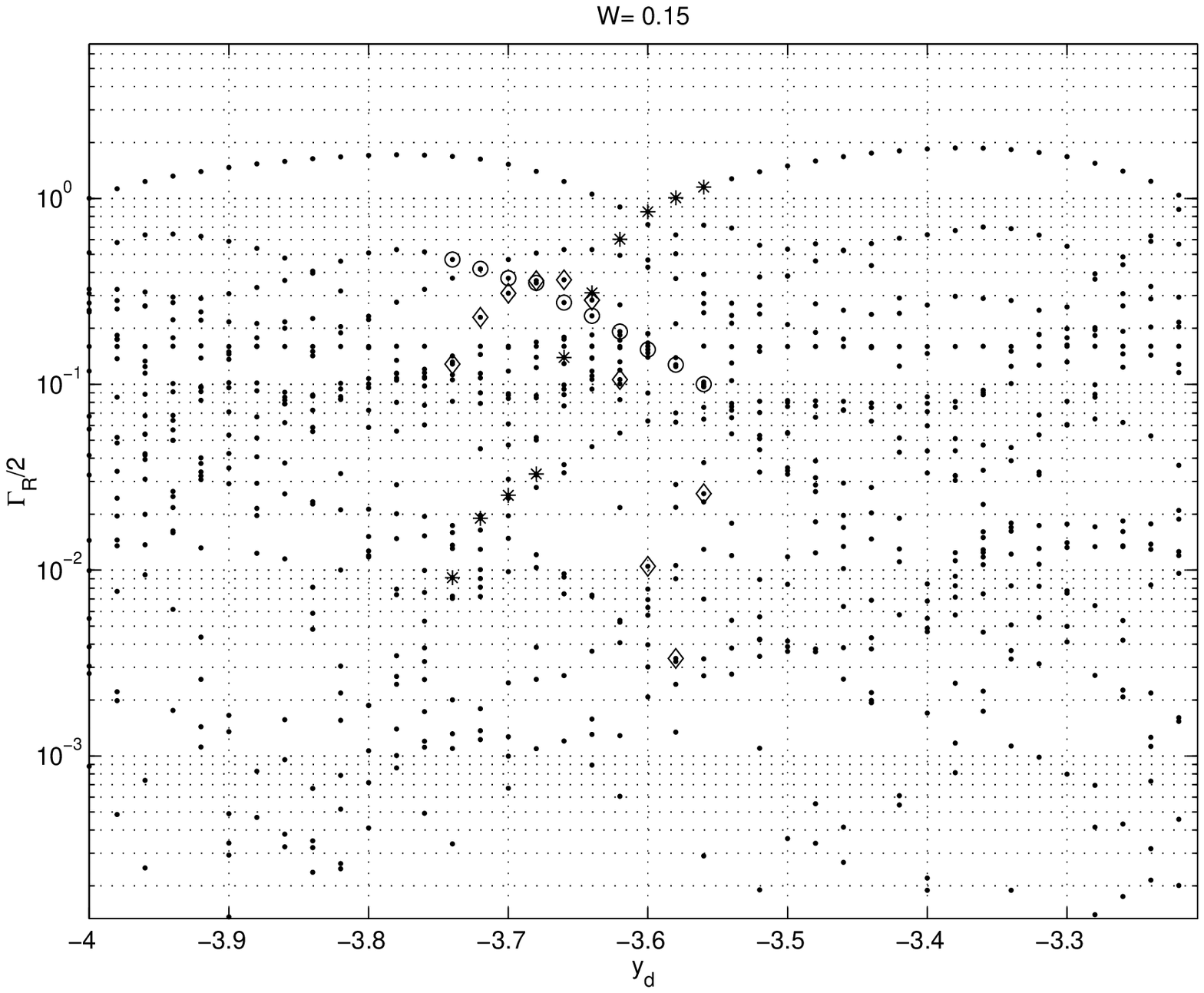,angle=0,width=12.5cm}
\end{minipage}
\caption{{\it Energies (top) and widths (bottom) as a function of $y_d$
for $w=0.15$ and $x_r$ = 1.5.
For 10 values of $y_d$, the poles of three
 states are marked by diamonds (A), stars (B) and circles (C). It is
  $E_A > E_B > E_C$
and $\Gamma_B > \Gamma_C > \Gamma_A $ at $y_d$ =
-3.56 while $E_A > E_C > E_B$ and
$\Gamma_C > \Gamma_A > \Gamma_B $ at $y_d$ = -3.74. 
The wavefunctions of these 3 states are shown in Figure 
\ref{fig:wf10y3}.
}}
\label{fig:eigen10y}
\end{figure}

The states whose 
wavefunctions are shown in Figure  \ref{fig:wf10y3} 
lie in the energy region around 24 where the coupling to the continuum 
is small. The mixing of the states  is 
varied by means of varying the area of the resonator.
In the upper part of the related Figure \ref{fig:eigen10y},  
we see typical avoided level crossings in the energies $E_R(y_d)$.
Here, the widths of the two states become comparable with one another.
This implies that $\Re(W)$ is decisive for the process.
In this case, the results are similar  to those known very well from
studies on closed systems with discrete states (see Section 3.1). 

We see, however, also the opposite case: the crossing
of the states $B$ and $C$ in Figure \ref{fig:eigen10y} is  free
along the real axis while the widths repel each other. In this case,
$\Re(W)$ is obviously small  
(the wavefunctions of the two states  are very different 
from one another, Figure \ref{fig:wf10y3}). Therefore,
 $\Im(W)$ is decisive, and the  levels can, 
according to Section 3.1, 
cross along the real axis.  

The variation of the widths of the resonance states
as a function of the coupling strength to the continuum is traced in
Figure \ref{fig:eigenw}. 
In each group of overlapping states, one collective state is formed 
whose structure  is determined by the channel wavefunction. This can be seen
very clearly by comparing the wavefunctions of the different collective states
which are similar to one another but have almost nothing in common with the
original wavefunctions of these states at small opening of the aperture 
(Figure  \ref{fig:wf3y}).
Here, the variation of the external mixing 
occurs mainly in the  $\Im(W)$:
the approaching of the states of a group in their positions as well as the 
trapping
of all but one state inside each group due to enlarging $\Im(W)$
 can be seen very clearly in Figure 
 \ref{fig:eigenw}.

It is interesting to compare Figure  \ref{fig:eigenw} with the results 
 for a slightly changed geometry of the cavity. In 
 \cite{seromupepi}, the disk is smaller and all states between the
two thresholds belong to one group.  According to this, 
only one broad state is formed at full opening of the slide.
 
The avoided crossing  of the two broad states in the lower part of
Figure  \ref{fig:eigenw}
occurs according
to the schematical picture with $i w_{\rm ex}$ (Figure  \ref{fig:eigenwf}.d, 
level attraction
and width bifurcation)  with the only difference that
not only the non-diagonal matrix elements of ${\cal H}$ depend on the coupling
strength determined by $w_{\rm ex}$, but also the  diagonal ones. 
This case is studied in detail analytically and
numerically in the framework of a schematical model in \cite{mudiisro}. 

In \cite{reichl2}, avoided level crossings 
in a closed resonator under the influence of a driving field are studied.
The results show avoided level crossings with and without permanent 
mixing of the wavefunctions
in a  similar manner as in the open resonator studied by us. 

The relation between the peaks in the conductance, the Wigner delay times 
and the 
positions of the states in the closed resonator is studied in \cite{reichl1}. 
The results of the present paper show that
the conductance peaks are related to the positions
of the resonance states in the {\it open} resonator.
The peaks are, generally, the result of 
interferences between the resonance states.  

Altogether, the interplay between $\Re(W)$ and $\Im(W)$ 
leads, as a rule, to permanent mixings of the wavefunctions. 
Level repulsion along the real axis is 
caused by $\Re(W)$ 
 while bifurcation of the widths (resonance trapping)   is  caused by 
$\Im(W)$. Both processes may interfere.
As a result, different time scales may appear and
the energy dependency of the conductance changes with the 
degree of opening of the system in a non-trivial manner.

\section{Conclusions}

The interaction $W$ of resonance states via the continuum of decay channels 
consists of the hermitian part  $\Re(W)$
 and the anti-hermitian part $\Im(W)$. Both terms have to be considered
 not only in a many-body system \cite{drozdz} but also in the micro-wave
 billiard as shown in the present paper. Some results show the 
 dominance of $\Im(W)$, others the dominance of $\Re(W)$. The avoided
 crossing of the resonance states in the complex plane may appear, under
 certain conditions, as a free crossing along the real axis or along the 
imaginary axis.

The interplay between the hermitian and anti-hermitian parts of the 
coupling operator between two resonance states via a common continuum 
may lead, in some cases, to a bifurcation of the widths (resonance trapping). 
In other cases, it may lead to a repulsion of the states
along the real energy axis.
The interaction $W$ 
introduces, as a rule, permanent changes in the wavefunctions of the
resonance states. Under certain conditions, the system may be stabilized
dynamically \cite{marost}.

The resonance picture of a microwave resonator shows all the 
characteristic features which are known from open quantum systems with 
two-body forces between the constituents. This result means that the 
interaction between the resonance states at the avoided level crossings
in the complex plane plays an important role for the mixing of the
wavefunctions.
As an example, the wavefunctions of the collective short-lived states are 
strongly mixed in the set of wavefunctions of the closed system. They are 
quite different from those of the original states at small coupling to the 
continuum. 

The statistical theory (random matrix theory) describes resonance states of
 an almost closed system. The poles of the $S$-matrix are near to the 
 real axis and $\Im(W)$ is small. The  
effective Hamilton operator is 
  ${\cal H} = {\cal H}_0 + \Re(W) + \Im(W) =
\Re({\cal H}) - i V V^\dagger $  where
 the $V$ are the coupling vectors  between 
discrete and scattering states \cite{mawe}. The level repulsion
along the real energy axis is embodied in $ \Re({\cal H})$ 
by choosing, e.g., the Gaussian orthogonal
 ensemble. 
Under these conditions, the effects caused by the interplay between
$\Re({\cal H})$ and  $\Im({\cal H})$ can be neglected to a 
good approximation. The 
results obtained in the present paper show, however, 
that the situation is different when the system is really open, i.e.
when $\Im({\cal H})$ and $\Re({\cal H})$ are of the
{\it same} order of magnitude.
In this case, the interplay between the two parts of ${\cal H}$
causes {\it non-negligible} effects
which are not considered in the statistical theory.  
The avoided crossing of resonance states in the complex plane embodies
 the interplay between   resonance trapping and 
level repulsion along the real axis.

\vspace{.8cm}

\noindent
{
{\bf Acknowledgment:} Valuable discussions with J. Burgd\"orfer,
M. M\"uller,
J. N\"ockel, K. Richter and M. Sieber 
are gratefully  acknowledged. 
}

\newpage

%\begin{figure}
%\nonumber
%\begin{minipage}[tl]{10cm}
%\psfig{file=wf1-bw.eps,angle=0,width=10cm}
%\end{minipage}
%\begin{minipage}[ml]{10cm}
%\psfig{file=wf3-bw.eps,angle=0,width=10cm}
%\end{minipage}
%\begin{minipage}[bl]{10cm}
%\psfig{file=wf5-bw.eps,angle=0,width=10cm}
%\end{minipage}
%\end{figure}

\begin{figure}
%\begin{minipage}[tr]{10cm}
%\psfig{file=wf7-bw.eps,angle=0,width=10cm}
%\end{minipage}
%\begin{minipage}[mr]{10cm}
%\psfig{file=wf9-bw.eps,angle=0,width=10cm}
%\end{minipage}
%\hspace*{1.9cm}
%\begin{minipage}[hr]{6.7cm}
%{\vspace*{.8cm}
\caption{\it 
The wavefunctions of the 3 states A (left), B (middle) and 
C (right) shown in Figure  
\ref{fig:eigen10y}
for $x_r$ = 1.5 and $y_d$ = -3.56 (1),  -3.58 (2) 
 -3.60 (3),  -3.62 (4), -3.64 (5), -3.66 (6), -3.68 (7),
 -3.70  (8), -3.72  (9), -3.74  (10). 
}
%}
%\vspace*{2.5cm}
%\end{minipage}
\label{fig:wf10y3}
\end{figure}

\begin{figure}
%\begin{minipage}[t]{15cm}
%\psfig{file=wf-br-bw.eps,angle=0,width=15cm}
%\end{minipage}
\caption{{\it  The wavefunctions of the 3 broad states shown in the lower part
of Figure \ref{fig:eigenw} at $w$ = 0.4 ({\rm a, b, c}) and $w$ = 0 
({\rm d, e, f}). The
state in the middle ({\rm b, e}) becomes trapped by the state to the left
({\rm a, d}).   }
}
\label{fig:wf3y}
\end{figure}

\end{document}